\documentclass[%
aip,jcp,amsmath,amssymb,
reprint,%
]{revtex4-1}

\usepackage[utf8]{inputenc}
\usepackage{mathtools}
\usepackage{amsmath}
\usepackage{graphicx}

\makeatletter
\newcommand*{\citenst}[2][]{%
  \begingroup
  \let\NAT@mbox=\mbox
  \let\@cite\NAT@citenum
  \let\NAT@space\NAT@spacechar
  \let\NAT@super@kern\relax
  \renewcommand\NAT@open{[}%
  \renewcommand\NAT@close{]}%
  \cite[#1]{#2}%
  \endgroup
}
\makeatother

\usepackage{xcolor}
\usepackage{todonotes}

\newcommand{\SI}{Supplemental Information}

\begin{document}
\title{Computational Study of Mechanochemical Activation in Nanostructured Triblock Copolymers}
\author{Zijian Huo}
\affiliation{Department of Chemistry, University of Pittsburgh, 219 Parkman Ave., Pittsburgh, PA, USA}
\author{Stephen J. Skala}
\affiliation{Materials Science and Engineering, Grainger College of Engineering, University of Illinois, Urbana-Champaign, IL 61801 }
\author{Lavinia Falck}
\affiliation{Department of Chemistry, University of Pittsburgh, 219 Parkman Ave., Pittsburgh, PA, USA}
\author{Jennifer E. Laaser}
\affiliation{Department of Chemistry, University of Pittsburgh, 219 Parkman Ave., Pittsburgh, PA, USA}
\author{Antonia Statt}
\email[email:]{statt@illinois.edu}
\affiliation{Materials Science and Engineering, Grainger College of Engineering, University of Illinois, Urbana-Champaign, IL 61801 }

\begin{abstract}
Force-driven chemical reactions have emerged as an attractive platform for diverse applications in polymeric materials. However, the network topologies necessary for efficiently transducing macroscopic forces to the molecular scale are not well-understood.
In this work, we use coarse-grained molecular dynamics simulations to investigate the impact of network topology on mechanochemical activation in a self-assembled triblock copolymers.  We find that mechanochemical activation during tensile deformation depends strongly on both the polymer composition and chain conformation in these materials, with activation requiring higher stress in materials with a higher glassy block content, and most activation occurring in the tie chains connecting different glassy domains. Our work suggests that changes in the network topology significantly impact mechanochemical activation efficiencies in these materials, suggesting that this area will be a fruitful avenue for further experimental research.

\end{abstract}

\keywords{mechanochemistry, block copolymers, coarse-grained simulation}

\maketitle
\date{\today}

\section{Introduction}
The ability to drive chemical reactions by applying force across chemical bonds has received significant attention as a platform for strain sensing, chemical catalysis, and self-healing properties\cite{Peterson2014,Li2015,Li2016,Piermattei2009,Groote2013,Diesendruck2013,Wang2015b}.
A key requirement for these applications is the ability to efficiently transmit force to the force-responsive bond.
Polymer chains can serve as effective ``handles'' for pulling on chemical bonds, both in solution\cite{Potisek2007,Su2018} and in polymer melts and crosslinked networks\cite{Davis2009,Wiggins_2013,Boulatov2017}.
However, the molecular weight of the polymer and the topology of the network strongly impact whether chains experience high enough forces to activate before the material relaxes and whether the forces are evenly distributed across all chains, which in turn determines the overall activation efficiency.

Understanding what features of the network topology most favor efficient activation is thus a critical knowledge gap that must be addressed to facilitate development of useful mechanochemical materials.
Block copolymers offer an attractive option for simultaneously controlling the network topology, and thus transmission of bulk forces to the molecular scale, and the overall mechanical properties of the material\cite{Bates1999,Bates2016,Inoue1971,Pakula1985}.
ABA triblock copolymers with long enough chain lengths ($N$) and interaction parameters ($\chi$) microphase segregate into ordered morphologies ranging from spherical to cylindrical to lamellar, depending on the volume fractions of each block type\cite{Matsen1999,Matsen2000}.
When the A blocks are glassy and the B blocks rubbery, the self-assembled A domains serve as physical crosslinks that enable the B chains to behave as an elastomeric network\cite{Inoue1971,Pakula1985,Tong2000}.
Importantly, the self-assembled structures not only have different overall symmetries, but also have different ratios of tie chains to loop chains\cite{Matsen1999}.
Because tie chains are expected to be more elastically-effective (and bear more force) than loop chains, changing the morphology of the material should drive substantial changes in the efficiency of mechanochemical activation in these materials\cite{Makke2012}.

To date, testing this hypothesis experimentally has been difficult.
In 2013, Jiang \emph{et al.} synthesized polystyrene-\emph{block}-poly(n-butyl acrylate)-\emph{block}-polystyrene triblock copolymers with a mechanochemically-active spiropyran unit in the middle of the rubber n-butyl acrylate block\cite{Jiang2013}.
While they did observe some dependence of activation on the weight fraction of the glassy polystyrene blocks, the relatively low molecular weights of the polymers and the low segregation strength of polystyrene and poly(n-butyl acrylate) precluded rigorous investigation of the role of morphology in mechanochemical activation.
Ramirez \emph{et al.} have also investigated mechanochemical activation in cast and electrospun triblock copolymers containing \emph{gem}-dibromocyclopropane mechanophores in the midblock\cite{Ramirez2014}.
They found that the triblock architecture promoted more activation than the midblock alone, but the inclusion of only a single polymer composition again made drawing conclusions about the role of morphology impossible.
While these studies are \emph{suggestive} of the role of network connectivity in determining the efficiency of mechanochemical activation in triblock copolymers, significant work remains to elucidate the role of specific topological features of the network in promoting or hindering mechanochemical activation in these materials.

Molecular dynamics simulations are a powerful way to investigate the physics of processes like mechanochemical activation in different chains within a polymer sample.
To date, theoretical treatments of mechanochemical activation have relied on Bell's model to calculate the probability of activation of a mechanophore from the force that it experiences\cite{Bell1978}. 
This model has been successfully applied in both continuum network models\cite{Silberstein2013,Silberstein2014,Wang2015a} and molecular dynamics simulations of randomly crosslinked networks, where it was found that the chains experiencing high enough forces to activate make up only a very small subset of the sample\cite{Adhikari2003}.
While force-based probabilistic models appear to qualitatively capture the trends in mechanochemical activation in real polymer materials, they cannot capture a number of features that are likely to be important in real chemical systems, including inhomogeneous environments, polymer chain conformations and network structure~\cite{Dubach2021}, and changes in mechanophore length upon activation\cite{Su2018}.
Thus, new models capable of directly capturing the force-driven transition in the state of the mechanophore will be necessary for fully elucidating the mechanisms of mechanochemical activation in complex systems such as triblock copolymers.

In this work, we address this challenge by developing a coarse-grained model for a simple mechanophore that can be incorporated into any bead-spring model of a polymeric material, and using it to investigate mechanochemical activation in triblock copolymers with different morphologies, as illustrated in Fig.~\ref{fig:schematic}. 
This model incorporates a bond double-well potential, which is inspired by prior work using pair potentials with two minima to model mechanochemical activation in an atomistic system\cite{Manivannan2016}.  The double-well potential has the advantage that the barrier height, activation length, and relative energies of the activated and inactivated states can be tuned arbitrarily, and the amount of activation in the material can be read out directly by monitoring the fraction of mechanophore units in the ``active'' potential well.
We find that activation is dependent on sample morphology, chain conformations and spatial location within the morphology.
In Section~\ref{sec:methods}, we will briefly introduce the model and simulation techniques used in this work, before discussing the results in Section~\ref{sec:results} and conclusions in Section~\ref{sec:conclusions}.

\section{Methods~\label{sec:methods}}

\subsection{Model~\label{sec:model}}
Molecular dynamics simulations of symmetric ABA triblock copolymers were carried out using a coarse-grained Kuhn segment level bead-spring model~\cite{Everaers2020}.
In this model, pair interactions are described by standard Lennard-Jones (LJ) potentials~\cite{Jones1924}, and bonds are modeled with finite extensible nonlinear elastic (FENE) potentials~\cite{Grest1990,Grest1986, Kremer1990}.
The overall chain length, $N$, and volume fraction of the glass A and rubbery B blocks ($f_A$ and $f_B$, repspectively) are controlled by changing the number of beads per chain, and the mechanical properties of each block and their interaction parameter, $\chi_{AB}$, are controlled via the interaction energies of the beads.

\begin{figure}[!h]
    \centering
    \includegraphics[width=8.6cm]{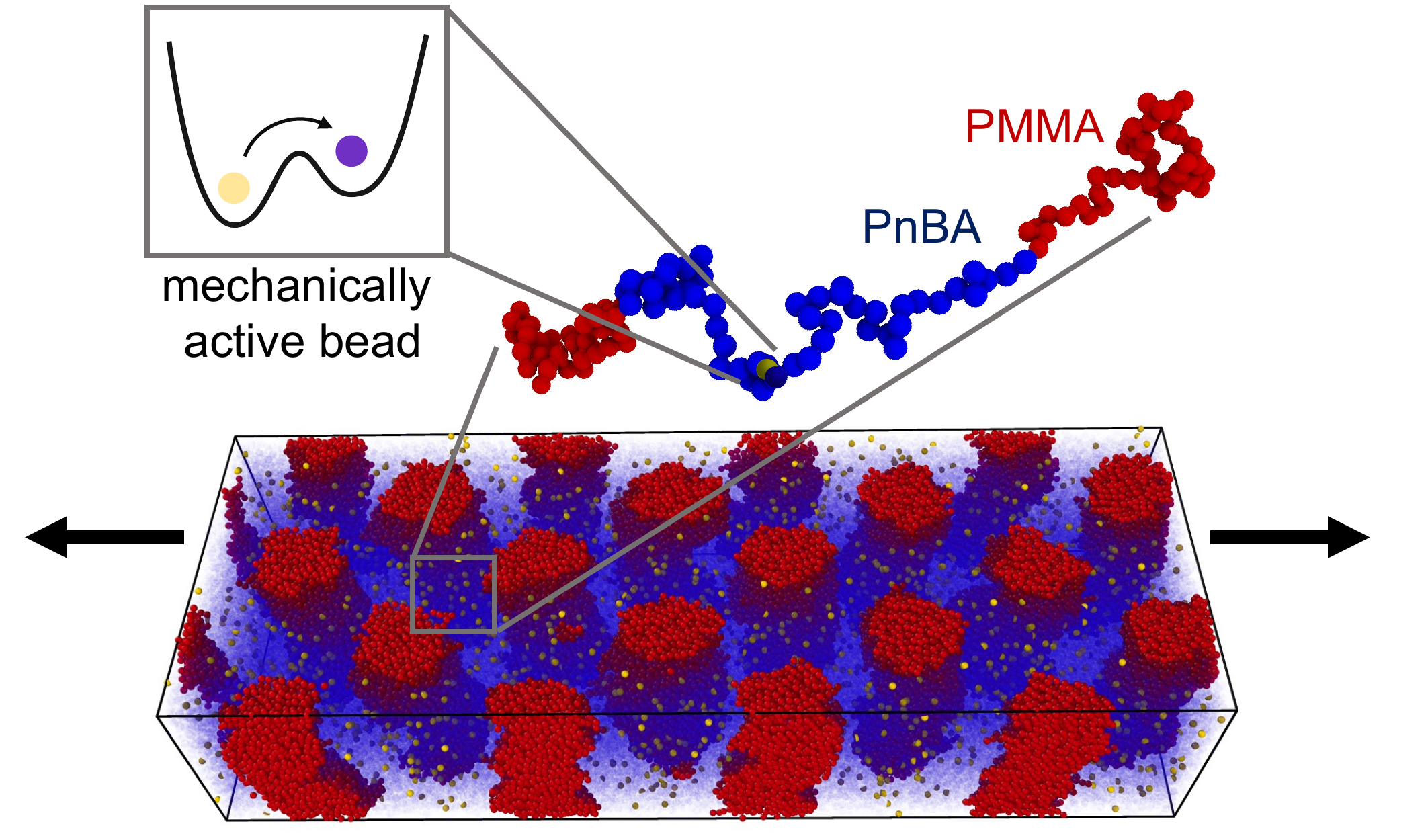}
    \caption{Schematic sketch of the system. Each triblock copolymer contains one mechanically-active bead (yellow or magenta) in the center of the rubbery PnBA midblock (blue), and two glassy PMMA endblocks (red). Activation of the mechanophore bead, which is modeled via a double-well potential, is monitored during uniaxial deformation of the microphase-separated sample (cylindrical morphology shown here).}
    \label{fig:schematic}
\end{figure}

In this work, we used the Kuhn segment molecular weights~\cite{Guo2006,fetters2007chain}, average amorphous densities~\cite{wypych2016handbook,fetters2007chain}, and glass transition temperatures~\cite{wypych2016handbook,Zabet2017} for both blocks to parameterize the coarse-grained model to describe a poly(methyl methacrylate)-\emph{block}-poly(n-butyl acrylate)-\emph{block}-poly(methyl  methacrylate) (PMMA-\emph{b}-PnBA-\emph{b}-PMMA) polymer system.  The PMMA-\emph{b}-PnBA-\emph{b}-PMMA system was chosen for this work because it forms well-ordered morphologies and can be synthesized via controlled radical chemistries that are compatible with the widely-studied spiropyran mechanophores\cite{Ruzette2006,Jiang2013}.
Thus, the index $A$ denotes PMMA Kuhn segments, and the index $B$ describes the PnBA Kuhn segments, with a block architecture as sketched in Fig.~\ref{fig:schematic}. 
The mass of an A bead (representing a PMMA Kuhn segment) was chosen to serve as unit of mass $m_A = 1\,m$, and the LJ diameter of the A particles $\sigma_{AA} = 1\,\sigma$ and their characteristic LJ interaction energy $\epsilon_{AA} = 1\,\epsilon$ were chosen as the units of length and energy. The derived units of temperature, pressure, and time were then $T^{*} = kT/\epsilon$, $p=\epsilon\sigma^{-3}$, and $\tau = \sigma\sqrt{m/\epsilon}$, respectively.
The parameters for the B beads (representing PnBA Kuhn segments) were then given by $m_{B} = 1.6\,m$, $\sigma_{BB} = 1.2\,\sigma$ and $\epsilon_{BB} = 0.6\,\epsilon$ (see \SI).
The cross-interaction parameters were finally chosen using Lorentz~\cite{Lorentz1881} and Fender-Halsey~\cite{Fender1962} mixing rules, yielding $\sigma_{AB} = 1.1\,\sigma$ and $\epsilon_{AB} = 0.75\,\epsilon$, where $\epsilon_{AB}= 0.75\,\epsilon$ ensures that the polymer melt phase separates under the investigated conditions. 

For computational efficiency, all LJ pair interactions were cut at $r_\text{cut}=3\sigma_{ij}$ and a smoothing function~\cite{Statt2020} was applied from  $r_\text{on}=2.5\sigma_{ij}$ to $r_\text{cut}$ to transition both energy and force smoothly to zero at $r_\text{cut}$. 

Experimental studies of mechanochemical activation typically use mechanophores such as spiropyran\cite{Davis2009} or naphthopyran\cite{Versaw2020}, which fluoresce or change color in response to a covalent bond breaking event. The optical signal can then be used to quantify the number of activated mechanophores in the material.  Because the coarse-grained simulations cannot capture this level of chemical detail, the mechanophore was instead simulated using a generic force activated bead with a double well bond potential (Fig. \ref{fig:schematic}, inspired by the Bell~\cite{Bell1978} and Cusp~\cite{Dudko2006} models and prior work in atomistic simulation systems\cite{Manivannan2016}. The first minimum corresponds to the deactivated state, and the second minimum represents the activated state. Changing the parameters of the double well potential allows the activation barrier between the two minima and the relative energies and positions of the minimia to be tuned arbitrarily. To avoid potential unphysical bond crossing in the activated state, we also included an additional mechanophore bead in the middle of the linear chains. This additional bead is held in place by two harmonic bonds to its adjacent $B$ beads, and a harmonic angular potential with a minimum at $180^\circ$. The details for the double well potential can be found in the \SI. 

All simulations were carried out in HOOMD-blue 2.9.2~\cite{Anderson2020,Phillips2011} and CUDA 10.1 on a single graphics processing unit (GPU) (NVIDIA GeForce GTX 1080Ti). The double well potential was implemented in a custom plugin~\cite{azplugins} for HOOMD-blue.

\subsection{Sample Generation/ Equilibration\label{sec:equilibration}}
To obtain equilibrated triblock copolymer samples with different morphologies, the volume fraction $f_A$ of the A block in the polymer chain model was varied from 0.16 to 0.50, targeting spherical, cylindrical, and lamellar phases as shown in Table~\ref{tab:systems}.




\begin{table}[!h]
\begin{center}
\caption{Polymer architectures and configurations \label{tab:systems}}
\begin{tabular}{llllll}
& \textbf{chain architecture} & morphology & $f_A$  & $V$ &  \\ \hline
& $A_{8}B_{50}A_{8}$ & spherical & 0.16 & $72\sigma \times 72\sigma \times 72\sigma$ &  \\
& $A_{15}B_{50}A_{15}$ & cylindrical & 0.26 & $80\sigma \times 69\sigma \times 40\sigma $&  \\
& $A_{43}B_{50}A_{43}$ & lamellae & 0.50 & $92\sigma \times 92\sigma \times 46\sigma$ &  
\end{tabular}
    \includegraphics[width=6cm]{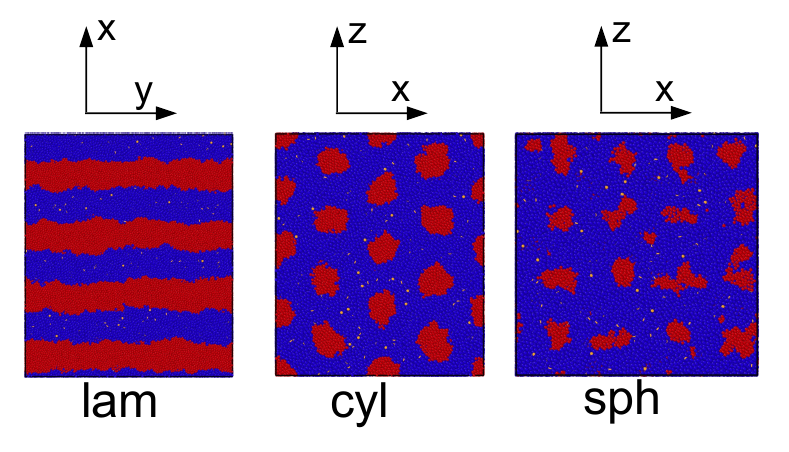}
\end{center}
\end{table}


 The polymer chains were placed randomly in a cuboid simulation box using standard periodic boundary conditions with the volume $V = L_{x} \times L_{y} \times L_{z}$ and the number density as $0.445\,\sigma^{-3}$, as shown in Table~\ref{tab:systems}. The randomly-placed polymer chains were then annealed toward the target microphase-segregated structures using dissipative particle dynamics (DPD)~\cite{Groot1997,Warren2017}, which enables fast phase separation\cite{Ryu2016,Li2013,Parker2014}. 

Even with the fast equilibration typical for DPD, however, it  takes considerable time for the systems to phase separate and anneal into the ordered morphologies. Therefore, only a small number of test systems were directly equilibrated using DPD, resulting in the expected morphologies after 
$\sim 1 \cdot 10^{4}\,\tau$ for lamellae, $\sim 5 \cdot 10^{4}\,\tau$ for cylindrical, and $\sim 6 \cdot 10^{4}\,\tau$ for spherical morphologies. For the remaining samples, an external field with domain spacings matching those obtained from the test systems was applied to direct the self-assembly and speed up the equilibration, based on the method discussed in Ref.~[\citenum{Padmanabhan2016}]. 
Full details of the equilibration and directed self-assembly methods are provided in the \SI. 

After self-assembly, the soft potentials used for the equilibration runs were  replaced with the LJ and FENE potentials described in the model section~\ref{sec:model}, and the timestep was consequently decreased to $\Delta t = 0.005\,\tau$ to ensure numerical stability and energy conservation. The external field was removed and further equilibration was performed in the NVT ensemble using a Langevin thermostat and a temperature  $T = 1.0\,\epsilon/k_{B}$ for $2.5 \cdot 10^{3}\,\tau$. 
After that, the equilibrated configuration was cooled down linearly within $4.5 \cdot 10^{3}\,\tau$ to the desired temperature of $T = 0.35\,\epsilon/k_{B}$ using the NPT ensemble with the MTK barostat-thermostat~\cite{Martyna1994} with coupling constants $\tau = 1.0$ and $\tau_{p} = 1.2$ at constant pressure $p = 0 \epsilon\sigma^{-3}$. This temperature is well below the glass transition temperature $T_{g}\approx0.43\,\epsilon/k_{B}$ of the $A$ blocks~\cite{Buchholz2002}, and above $T_g$ of the $B$ blocks. 
The system was then further equilibrated at a constant temperature $T = 0.35\,\epsilon/k_{B}$ for another $500\,\tau$ to ensure an equilibrated morphology before the spiropyran units were incorporated into the center of the B beads. More information on the equilibriation steps can be found in the SI.

\subsection{Deformation\label{sec:deformation}}

The tensile response of the samples was then simulated by deforming the equilibrated configurations at a strain rate of $4 \cdot 10^{-4}\,\tau^{-1}$ in a NVT ensemble using a Langevin thermostat with a constant temperature $T = 0.35\,\epsilon/k_{B}$. The average stress $\Sigma$ was obtained from the simulation using the pressure tensor components, $\Sigma = \frac{3(P_{ii}-P)}{2}$ where $i$ is the  direction of elongation and $P$ the pressure, $P = \frac{\sum_{i}P_{ii}}{3}$ \cite{Rottach2004}.  All uniaxial deformations were repeated ten times with independently equilibrated samples. 

In previous studies of rubbery systems~\cite{Aoyagi2002} it was found that below a percent elongation of 150\%, affine deformation of chains takes place, and that between 150\% and 350\% morphologies are maintained, but chain deformation is no longer affine. Here, a maximum deformation of 400\% was used to ensure that the entire range of sample responses was investigated.

Similar to prior studies, in the lamellar morphology, some buckling~\cite{Makke2013} and cavitation~\cite{Makke2011} events were observed in the rubbery phase around a percent elongation of 200\%. The cylindrical and spherical samples showed deformation of the micro-domains but no large scale reorientation up to strains of approximately 170\% for the cylindrical and 70\% for the spherical samples.  Cavitation was only observed in the cylidrical samples toward the very end of the deformation, at strains above approximately 350\%.

\section{Results \label{sec:results}}
\subsection{Equilibrium properties}

Using the method described in Section~\ref{sec:equilibration}, ten independent replicas were equilibrated for each composition listed in Table \ref{tab:systems}.  All systems self-assembled into the morphologies expected from SCFT phase diagrams~\cite{Matsen1999,Matsen2000}.  
The average spacing of the morphologies were determined to be $\sim 18\sigma$ for the distance between $A$-rich spheres in the spherical morphology, $\sim20\sigma$ for the average distance between cylinders in the cylindrical morphology, and $\sim 26\sigma$ for the average spacing between lamellae, in agreement with SCFT predictions.
Agreement with SCFT predictions was further assessed by determining the fraction of loop and tie chains in each sample~\cite{Matsen1999}, where loop chains are chains whose ends are both in the same $A$ glassy domain, and tie chains are chains whose ends connect two different $A$ domains.
The results were in good agreement with SCFT predictions~\cite{Matsen1999,Aoyagi2002}, with tie-chain fractions of $0.73\pm0.01$, $0.62\pm0.04$ and $0.38\pm 0.03$ in the spherical, cylindrical, and lamellar morphologies, respectively.
Loop chains were further classified as ``hooked'' if they were entangled, or hooked onto, a loop chain originating from a different glassy domain. Details of the chain classification algorithm are provided in the \SI. 
On average, hooked chains were found to have 3 kinks in their primitive paths, whereas un-hooked loop chains had 1-2 kinks and tie chains typically had one. All morphologies were well phase-separated and therefore had a negligible fraction of dangling/cilia ends in their initial equilibrated states, with approximately 1.5\% dangling chains in the spherical sample as the highest observed quantity.  

This analysis indicates that the equilibration procedure described in Section \ref{sec:equilibration} successfully generated correctly equilibrated triblock copolymer samples, with the expected network topologies.
While some discrepancies in the domain spacings and loop and tie fractions were observed (see SI), such discrepancies are expected because the SCFT results are only valid for identical monomer sizes, whereas here the $B$ beads are slightly bigger than the $A$ beads.\cite{Matsen1999} SCFT results also show a weak dependence of the tie fraction with degree of segregation and copolymer composition, which are neglected in this comparison.\cite{Matsen1999}
These effects are expected to be small, however, and the overall agreement between the simulation results and the SCFT predictions supports that the samples investigated here are equilibrated properly.

\subsection{Stress strain behavior}

\begin{figure}[!h]
    \centering
      \includegraphics[width=8.5cm]{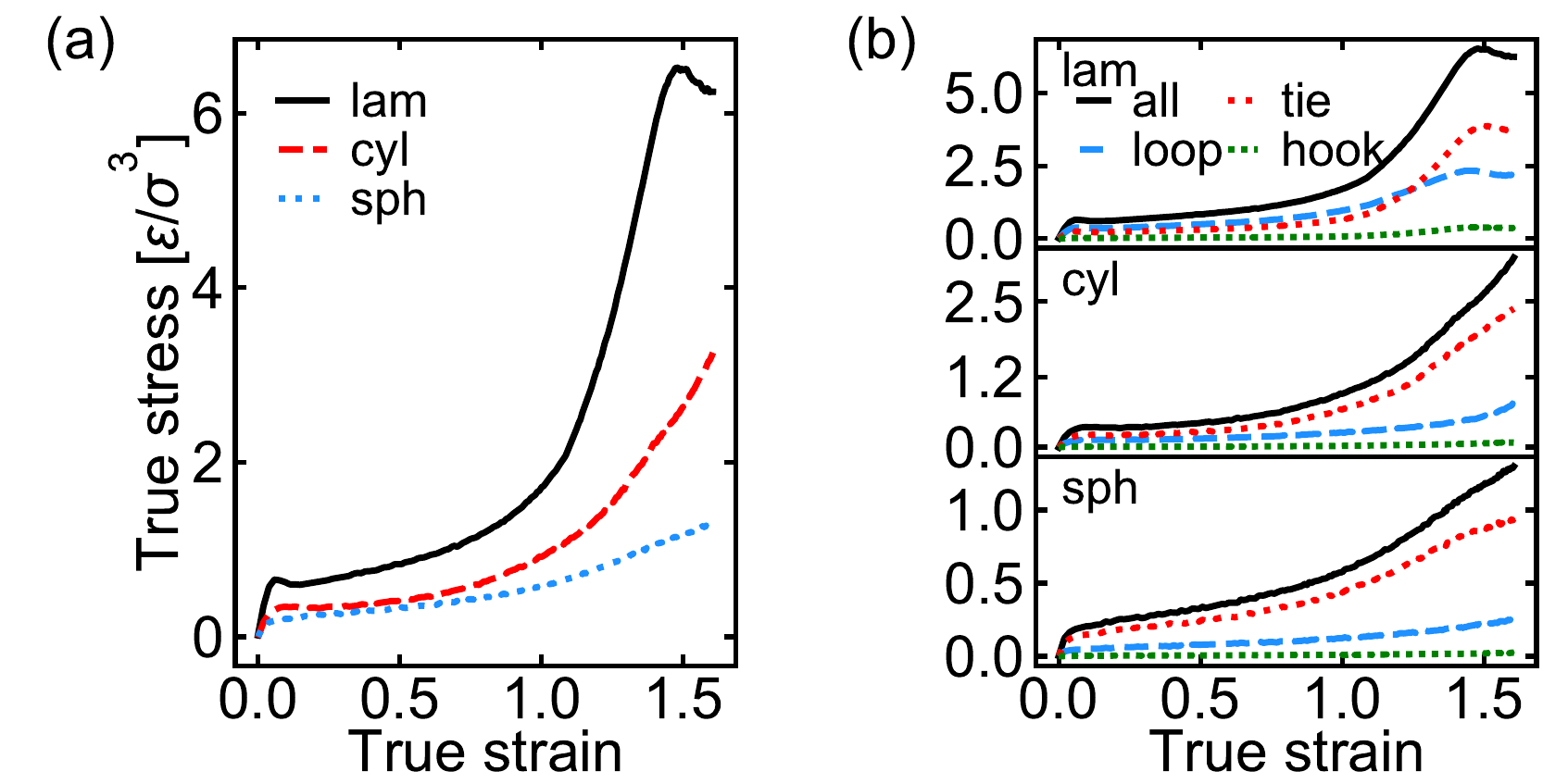}
    \caption{(a) Stress-strain curves for the lamellar (lam), cylindrical (cyl), and spherical (sph) morphologies and (b) contribution of each chain type  to the overall stress-strain curve. The contributions are normalized such that the addition of all curves results in the total stress on the sample. Each stress-strain curve reflects an average over ten independent simulation runs.}
    \label{fig:stress_strain_all}
\end{figure}

To investigate the stress-strain behavior as a function of morphology, the lamellar, cylindrical and spherical samples were deformed perpendicular to their periodic microstructure.  Samples were deformed up to 400\% elongation, corresponding to true strains between 0 and 1.63, and each stress-strain curve was averaged over ten independent simulation runs. The resulting overall stress-strain curves are shown in Fig.~\ref{fig:stress_strain_all}a. 
Overall, the stress-strain curves follow the expected behavior, with the stress increasing with the fraction of the glassy $A$ component in the sample. 
Consistent with prior results from simulations~\cite{Makke2012,Makke2013} and experiments~\cite{Adhikari2003,Tomita2017}, the absolute stress is highest for the lamellar morphology, followed by cylindrical morphology, with the spherical morphology having the lowest stress.

The stress-strain behavior exhibited a very short linear elastic regime at low strains, followed by a weak yield point for the lamellar sample, and a continuous increase of stress, indicating strain hardening at higher strain values. 
Generally, the behavior of the simulated BCP was somewhat more ``ductile'' than an experimental system would be. 
This effect is a known shortcoming of the LJ potential, which does not reproduce brittle failure modes under uniaxial deformation accurately.~\cite{Lin2019,Makke2011} In addition to the pair-potential, periodic boundaries stabilize plastic flow~\cite{Lin2019,Shi2010} and suppress crazing and shear banding. 
While we focus here on the activation behavior of the embedded mechanophore in the rubbery block and do not investigate failure modes associated with the glassy domains directly, detailed investigation of mechanophore activation in relation to material failure may be an interesting direction for future work.
    
The stress-strain behavior as a function of chain type is shown in Fig.\ref{fig:stress_strain_all}(b).
As expected, the tie chains carried the majority of the total stress~\cite{Makke2012}, especially for the cylindrical and spherical morphologies. 
In the lamellar sample, where tie chains make up 38\% of the sample, loop chains transmit roughly half of the stress, up to a strain of approximately 1.2. 
When normalized by the absolute number of each type of chain (see \SI), the average stress per chain is the same on all chains for low strain, up to a strain of roughly 0.8; for higher strains, the loop chains carry a lower than average stress, in contrast to tie and hooked chains, which carry higher than average stress.  
These results indicate that the contributions from the different types of chains obtained from the coarse-grained simulations are reasonable, and provide a sound basis for analyzing mechanochemical activation across different types of chains.

\subsection{Activation behavior}

\begin{figure}[!h]
    \centering
    \includegraphics[width=8.5cm]{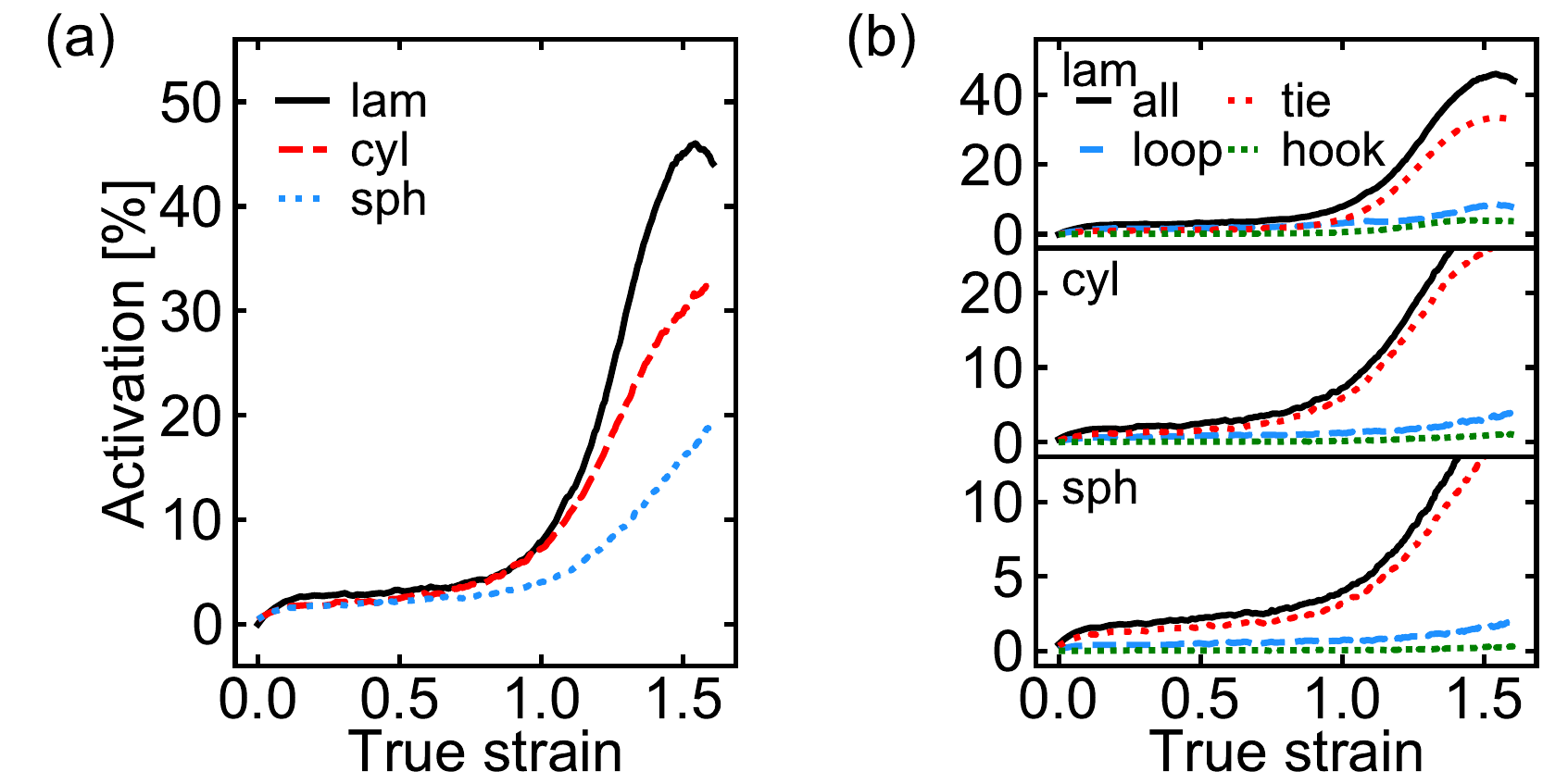}
    \caption{(a) Total activation of the mechanically active bond in different morphologies as function of strain, and (b) contribution of each chain type to the total activation for each morphology. Curves in (b) are normalized such that the addition of the activation for each individual chain type equals the total activation (black line).
    \label{fig:activation_stress_strain}}

\end{figure}

To quantify the strain-induced mechanochemical activation of each sample, the number of activated mechanophore beads was defined as the number of double-well mechanophore bonds in the second minimum, which was calculated using a simple distance-based cut-off criterion (where the cutoff was located at the local maximum of the potential between the two minima). 
The total activation measured for each morphology is presented in Fig. \ref{fig:activation_stress_strain}a.
At low strains, all morphologies had a very similar low level of activation of about 5\%, which is above the typically observed thermal activation of less than 2\%. (sph: (1.4 $\pm$ 0.2)\%; cyl: (1.8 $\pm$ 0.4)\%; lam: (1.8 $\pm$ 0.3)\%)
At high strains, the fraction of mechanophore units activated increased with increasing glassy block content, consistent with the higher stress required to strain these materials. 
Overall, peak activations of approximately 45 \% (lam), 30\% (cyl) and 20\% (sph) were observed at the highest true strain value of 1.6. 

As with the stress, the activation was also analyzed as a function of chain type.
As seen in fig.~\ref{fig:activation_stress_strain}(b), across morphologies, the vast majority of the activation occurred in the tie chains - even in the lamellar morphology, where the tie chains make up only 38\% of the sample and loop chains bear a comparable fraction of the total stress.
Loop chains showed significantly lower activation, even when normalized for their relative fractions (see \SI). 
This result suggests that control over the tie chain fraction is critical for optimization of mechanochemical activation in block copolymre networks.

The stress on individual chains, $\Sigma(\alpha)$, was calculated using the virial formula~\cite{Mott1992},
\begin{align}
 \Sigma_{ab}(\alpha)=-\frac{1}{V_\alpha}\sum_{i=1}^{N_\alpha}\left[m^{(i)}v_{a}^{(i)}v_{b}^{(i)}+{\frac {1}{2}}\sum _{j=1,j\neq i}^{N,N_\alpha}{\vec {r}}^{(ij)}_{a}{\vec {f}}_{b}^{(ij)}\right]
 \label{eq:stress_per_chain}
\end{align}
Here, $V_\alpha = \sum_{i=1}^{N_\alpha} V^\text{voro}_i$ is the volume of the chain, as determined by the sum of all individual bead Voronoi volumes of that chain~\cite{Mott1992,Makke2011}. This volume definition ensures overall correct normalization, i.e.,\ adding up all individual stresses on each chain in a system results in the total stress as determined by the pressure tensor components for the whole system. The second sum in Eq.~(\ref{eq:stress_per_chain}) runs over all beads in the chain of interest and over all particles in the system, such that all non-zero forces on a particular chain are counted. 

\begin{figure}[!h]
  \centering
    \includegraphics[width=8.5cm]{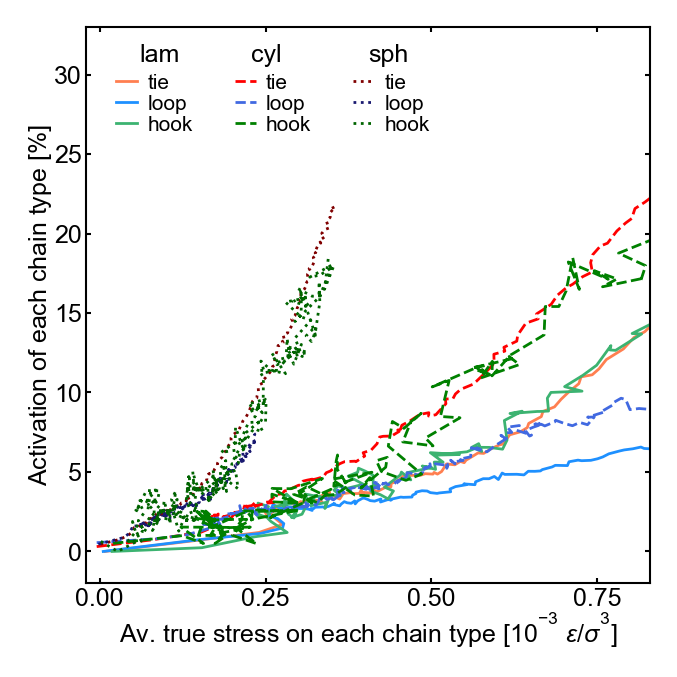} 
    \caption{Activation of each chain type as function of true stress on each chain chain type, for all three morphologies (lamellar - solid lines, cylindical - dashed lines, spherical - dotted lines) as indicated in the legend.
    \label{fig:tie_loop_activation_stress}}
\end{figure}

When activation is plotted as function of average stress on each chain instead of total true strain, as shown in Fig.~\ref{fig:tie_loop_activation_stress}, differences in the behavior of each morphology becomes apparent. Interestingly, the spherical morphology, with the lowest fraction of glassy block, had the highest activation at low stress values. The lamellar morphology, which had the highest activation as function of true strain (Fig.~\ref{fig:activation_stress_strain}), had the \emph{lowest} activation as function of stress on each chain type. 
From a materials design standpoint, this result means that a large fraction of tie chains or hooked chains is desirable when a high overall activation is desired, especially at large strain. If a response at low absolute stress values is desired, spherical morphologies should give the highest activation at relatively low stress.

As seen in Fig.~\ref{fig:tie_loop_activation_stress}, the hooked loop chains followed the tie chain activation behavior quite closely, responding to stress in a similar way. From a geometric perspective, this is expected, because two hooked chains should transmit forces very similarly to how a tie chain would~\cite{Makke2012}. When normalized by number of chains, i.e, average activation per chain type (see \SI), it becomes apparent that as the $f_A$ decreases, the average activation of the hooked chains also decreases slightly: in the lamellar morphology, hooked chains had a slightly higher average activation than tie chains; in the cylindrical sample they were about the same; and in the spherical morphology the hooked chains had a lower average activation than the tie chains.  This trend was consistent with the amount of hooked chains in each, with approximately 5\% hooked chains in the lamellar samples, 1.7\% in the cylindrial and 1.3\% in the spherical moprphology. 
In contrast, the loop chains showed a consistently lower activation at the same average stress per chain, which was especially pronounced in the cylindrical and lamella morphology.

\subsection{Chain conformations}

To gain more insight into details of the chain conformation in the rubbery $B$ region, primitive path analysis (PPA) was carried out on the entire deformation trajectory.~\cite{Karayiannis2009,Shanbhag2007,Kroger2005,Foteinopoulou2006}
In this analysis, the $A$ beads were held fixed, and the PPA was applied only onto the $B$ beads. 
The primitive path length $L_{c,B}$ and the end-to-end distance $R_{e,B}$ of the rubbery mid-block, as well as the number of kinks on the rubbery mid-block, were then extracted from the primitive path network.
Details of the PPA analysis are provided in the \SI.

\begin{figure}[!h]
    \centering
   \includegraphics{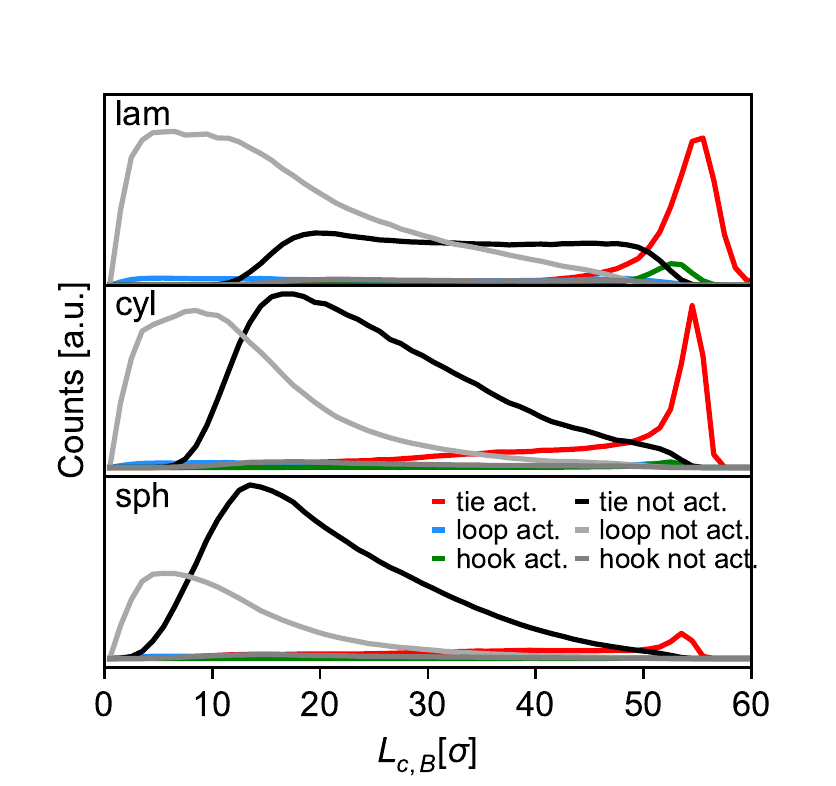}
    \caption{Contour length $L_{c,B}$ of the primitive path of the rubbery $B$ block for different chain types (activated: red, blue, green or not activated: black, grey, dark grey), as indicated.}
    \label{fig:ppa_length}
\end{figure}

Histograms of the primitive path contour lengths for all samples are shown in Fig.~\ref{fig:ppa_length}.
As seen in this figure, activated tie and hooked chains typically had a longer primitive path length than their non-activated counterparts. The activated tie chains have $L_{c,B}$  lengths around $55\sigma$, very close to maximum extension ($\sigma_{BB}N \approx 60\sigma$).
The loop chains did not show such a trend, with roughly the same fraction activated regardless of their $L_{c,B}$.  
Interestingly, the activation depended somewhat on the number of kinks in the chains: at high strain rates, activated tie chains had a slightly lower average number of kinks than the non-activated tie chains, and activated loop chains had a slightly higher average number kinks than the non-activated loop chains. Details of this analysis are give in the \SI.
Detailed analysis of the behavior of further sub-classes of chains (e.g.,\ loop chains entangled or hooked with tie chains, etc.) is beyond the scope of the present work, but may be an interesting avenue for future analysis.

\subsection{Spatial Localization of Activation/Stress}

Because the composition of self-assembled block copolymers varies spatially within the material, it is anticipated that mechanochemical activation may vary spatially within the $B$ regions, as well.  To investigate this problem, each of the ten independent runs for the lamellar and cylindrical morphologies were aligned and averaged to obtain spatial 2D maps of the fraction of $B$ beads, the local lateral stress, and the local fraction of activated SP beads. Samples were aligned by shifting the closest center of mass of a domain to the origin, and wrapping the shifted coordinates with respect to the periodic boundary conditions. The analysis was carried out at a true strain value of 1.2, or 300\% elongation, to capture the effect of deformation while the morphology domain patterns were still relatively intact. The spherical system was omitted from this analysis because the spherical $A$ regions became too distorted during elongation to allow robust alignment.

\begin{figure}[!h]
  \centering
       \includegraphics{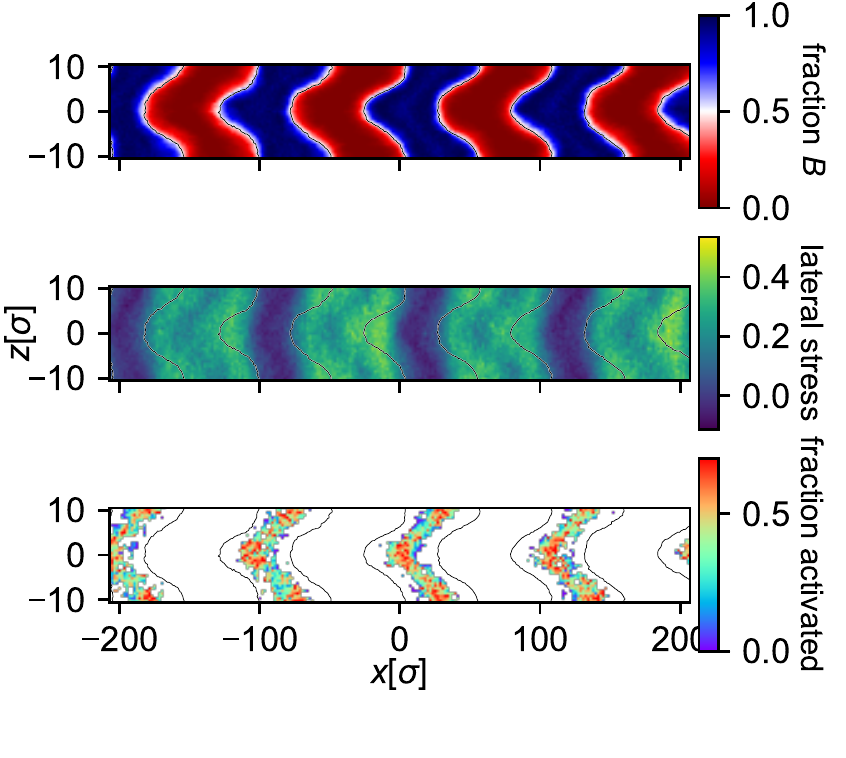} 
    \includegraphics{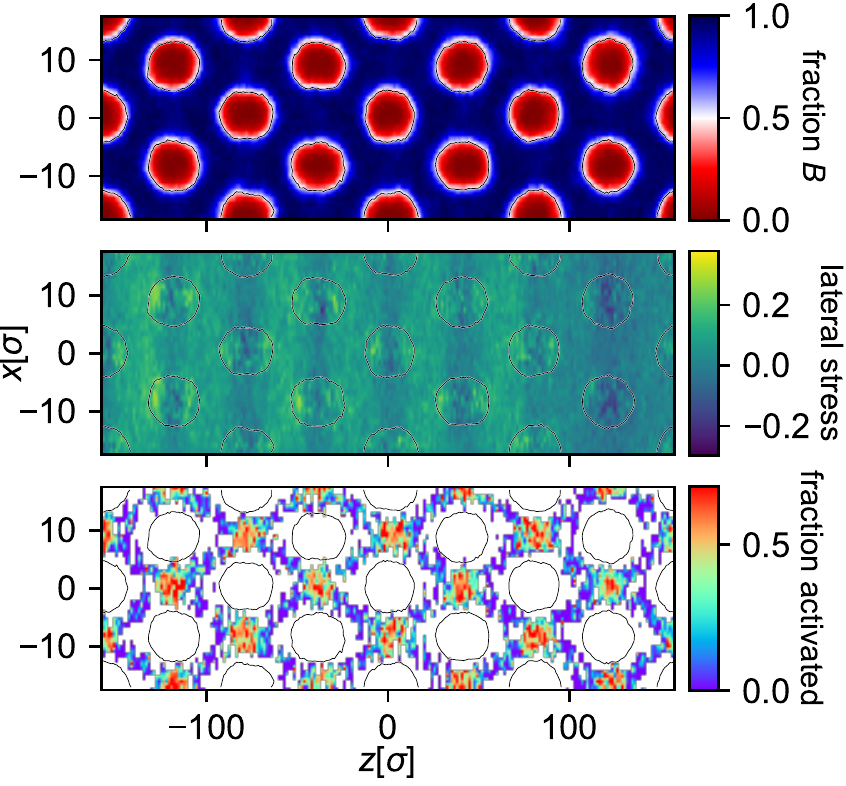}
    \caption{Spatial maps of the lamella (top) and cylindrical sample(bottom) at approx. 1.2 true strain.. Top panel shows the density, the middle the spatial localization of SP activation, and the bottom the local lateral stress. The bin size for each 2D histogram was $1\sigma$.}
    \label{fig:2D_density_activation_stress}
\end{figure}

As shown in the top panel of Fig. \ref{fig:2D_density_activation_stress}, the lamellar morphology underwent buckling and started to exhibit a chevron-like pattern at moderate to high strain. This behavior is consistent with prior observations in literature~\cite{Makke2013,Makke2012} for similar model triblocks. The buckling is caused by a Poisson effect due to the lateral stress mismatch in the glassy and rubbery phase.~\cite{Makke680} 
The exact modes of deformation and yield are expected to depend on strain rate and sample size, as shown in Ref.~[\citenum{Makke680}], although this effect was not investigated in this work.  
Negative lateral stresses emerged in the glassy domains, and positive lateral stresses in the rubbery regions, indicating some compression of the glassy domains and expansion of the rubbery domains under tension. Additionally, some localization of extreme values of the stress near the interfaces was apparent in the rubbery regions. 

Interestingly, activation in the lamellar samples was strongly correlated with the shape of the domains, and was localized towards the center and the tips of the chevron pattern, e.g., red areas in Fig.~\ref{fig:2D_density_activation_stress}. Note that the 2D activation map in this figure is normalized by the density of mechanophore beads, and regions with no mechanophore beads are indicated in white. Therefore, the observed pattern is not an effect of increased density of mechanophore units, which is also higher towards the center of the rubbery regions. 
The mechanophore activation appears to not be correlated with extreme values in lateral stress, but instead follows the pattern of domain deformation. 

Similarly, in the cylindrical morphology, a pronounced cross-hatch pattern appeared visible (lower panels in Fig.~\ref{fig:2D_density_activation_stress}), with the highest activation observed toward the middles of the rubbery regions. Again, extreme values of stress were located near the interface between the two phases. In general, in comparison to the lamellar morphology, the lateral stress in the cylindrical morphology was lower and had less pronounced positive or negative extremes.
The spherical morphology did not show any pronounced spatial activation features.  We note, however, that a two-dimensional projection of the spherical morphology leads to averaging of both rubbery and glassy domains, which might hide spatial features. In addition, the spherical morphology preserves less of its spatial order when deformed, making aligning the ten samples for averaging more difficult. 

\subsection{Orientation/Direction dependence}

\begin{figure}[!h]
  \centering
    \includegraphics[width=8.5cm]{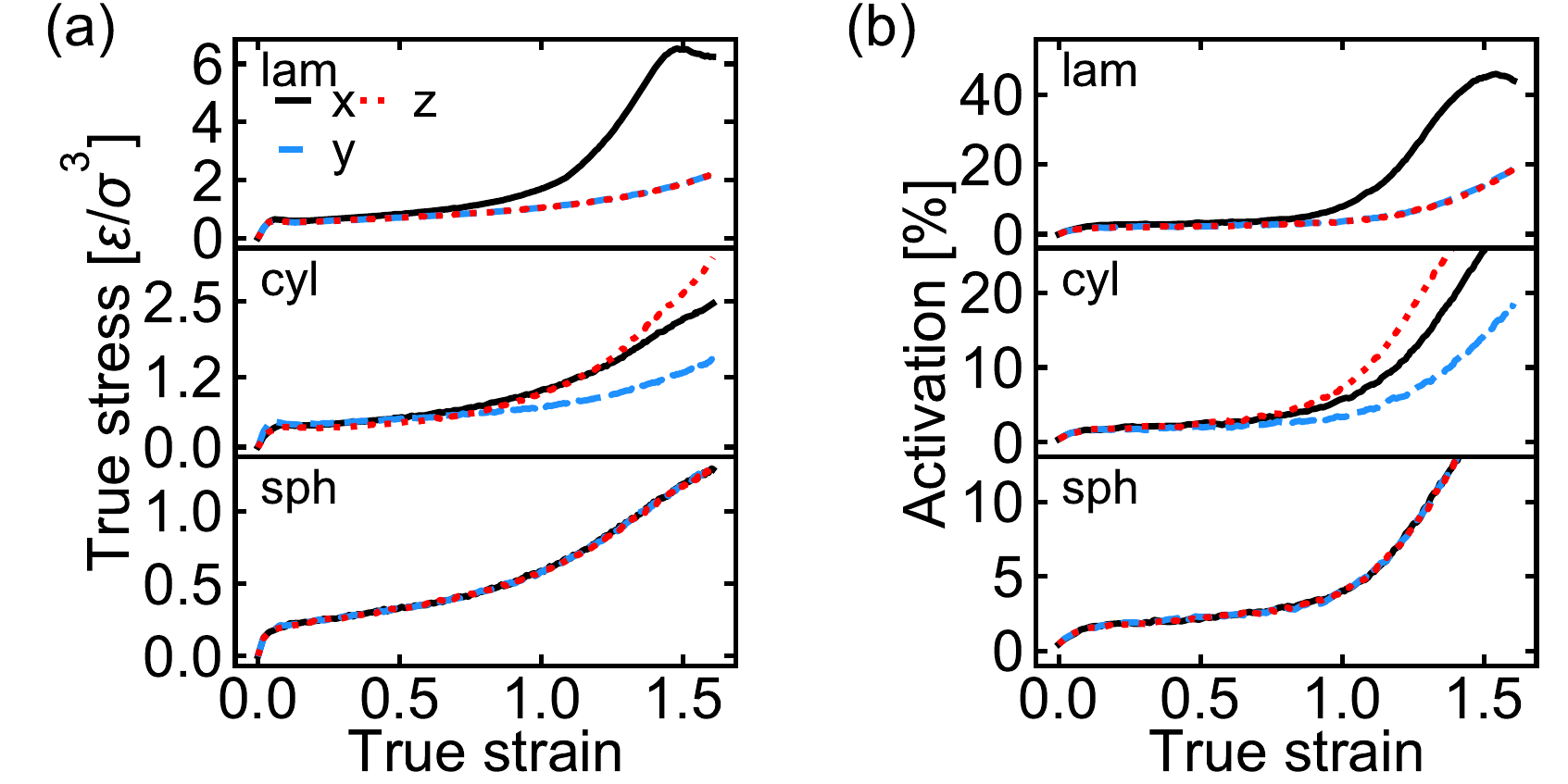}
    \caption{(a) Overall stress strain and (b) overall activation curves of each morphology, deformed in different directions. See Table~\ref{tab:systems} for snapshots with indication of the directions. $x$ is perpendicular to the lamellae, $z$ is along the cylinder hexagonal axis, $y$ is parallel to the cylinder main axis.}
    \label{fig:stress_directions}
\end{figure}

The preceding results are all for deformation of each morphology perpendicular to its characteristic domain pattern. Because real block copolymer samples contain a mixture of domain orientations, however, it is also important to understand how the morphology's orientation relative to the applied force affects mechanochemical activation.  To this end, additional uniaxial deformation tests were performed on each system, with force applied in both perpendicular directions.  
The direction-dependent stress-strain and activation-strain behaviors are summarized in Fig. \ref{fig:stress_directions}.
In the lamellar morphology, deformation in the $x$ axis reflects deformation perpendicular to the lamellae, while deformations in the $y$ and $z$ directions are both deformations parallel to the lamellae.  As expected, from this symmetry, the $y$ and $z$ responses were identical, and both the stress and activation were significantly lower for these deformations than for deformation perpendicular to the lamellae, where the chains are better aligned with the pulling direction.
In the spherical case, all three directions are equivalent and exhibited identical stress and activation responses, as expected from the symmetry of this morphology. In the cylindrical case, on the other hand, all three directions behaved sightly differently; in this case, the $x$ and $z$ deformations are both perpendicular to the cylinder axes, but because the $x$ axis is tilted $60^\circ$ from the axis of hexagonal symmetry, the deformation is not as well aligned with the chains connecting the glassy domains. The lowest stress was observed for deformations parallel to the long axis of the cylinders which is in agreement with prior experimental results.~\cite{Honeker2000,Zhang2019}
When the morphologies were deformed parallel to its characteristic domain direction, we did not observe any distinctive spatial patterns in activation, in contrast to the results shown on Fig. \ref{fig:2D_density_activation_stress}.

Overall, the strain behavior is consistent with that observed in tensile experiments on aligned block copolymer samples~\cite{Zheng2016,Wang2016,Honeker1996}, and the overall activation tracked the trend of the stress-strain curves.
Importantly, the deformation axis most perpendicular to the morphology domain always led to the largest activation; in a macroscopic sample with imperfectly aligned domains, one would observe some averaged overall activation and stress. 

\section{Discussion \label{sec:discussion}}

In this work, we utilized coarse-grained molecular dynamics simulations with a model mechanophore to investigate mechanochemical activation in a series of well-ordered block copolymers.  A key advantage of this approach, as illustrated in this work, is that it allows extraction of detailed information about how individual populations of chains do or do not participate in the activation process.   
As we show, modelling activation via molecular dynamics simulations allows the responses of tie, loop, and hooked chains to be analyzed separately, which is impossible in experimental systems.
We find that tie and hooked chains activate much more efficiently than loop chains, which confirms our original hypothesis and highlights the critical importance of network topology in determining the transmission of force to (and activation of) individual polymer chains.

These results suggest that block copolymer systems merit further experimental investigation as platforms for mechanochemical activation.  As shown in this work, changing the composition of the block copolymer (and its self-assembled morphology) impacts both the modulus of the material and the efficiency with which it activates a mechanophore at the center of the rubbery midblock.
For an application requiring activation at low stress and low strain, a spherical morphology may offer the best balance of low modulus and activation efficiency, while for an application requiring activation at high stress and high strain, a lamellar morphology will be more appropriate.

In this context, understanding the similarities and differences between the experimental and computational systems is critical.
First, the strain rate used in the computational work is orders of magnitude higher than those typically used in experimental measurements.  As a result, the computational model exhibits behaviors, such as cavitation, that likely do not play a significant role in low strain-rate experiments.  As shown in the Supporting Information, we find that lowering the strain rate drives slightly earlier activation of the mechanophore units as force the chains are able to relax and transmit force uniformly to the mechanically-active bond.  While this effect should not change any of the trends observed in terms of which chains activate and which do not, experimental systems may exhibit some differences in absolute activation profiles due to more efficient redistribution of force on the experimental timescale.

Second, the morphologies used in this work were well-aligned relative to the stretching direction, while in experimental systems, the domain orientations are typically randomized, with grains on the order of 100s of nm.  To a first approximation, measurements on experimental systems are thus expected to give a result that is essentially the average of the different orientations.  In spherical systems, this effect should not change the overall activation profiles, because the spherical morphology is already symmetric along all three axes.  In lamellar systems, on the other hand, this effect might significantly lower the average activation.
This result may be further complicated by reorientation of domains under tensile deformation\cite{Honeker1996}.
Finally, in experimental systems, the oriented grains have edges.  As shown in Fig. \ref{fig:2D_density_activation_stress}, activation in lamellar samples that buckle is localized near the tips of the buckled chevron pattern.  This result suggests that activation may also be more pronounced at grain boundaries, where different orientations of the lamellar morphology join together.
Understanding the effects of domain orientation and domain boundaries will be critical for understanding experimental mechanochemical activation profiles in these systems.

Finally, beyond the block copolymer system investigated in this work, we believe that explicit modeling of activation in coarse-grained simulations has the potential to expand understanding of a wide range of other mechanochemically-active polymer systems as well.
Loops, which our work shows are elastically and mechanochemically mostly inactive in the block copolymer systems, also play an important role in the mechanical properties of regularly-crosslinked networks\cite{Zhong2016}, and controlling loop defect content may offer new opportunities to improve activation in these types of networks.
By directly modelling the mechanophore rather than inferring activation rates from a measured force, we also open the possibility of modelling mechanical responses of systems incorporating scissile mechanophores that cleave the polymer chains upon activation\cite{Li2015,Klein2020}, or mechanophores that change the length of the polymer chains as they are activated\cite{Su2018}.  
Coarse-grained MD simulations should thus be an attractive tool for investigation of a wide variety of mechanochemical phenomena.

\section{Conclusions \label{sec:conclusions}}

In this work, we developed a simple model mechanophore that can be incorporated into coarse-grained bead-spring models of polymeric materials, and used it to investigate mechanochemical activation in a series of well-ordered triblock copolymers.
The coarse-grained model for the polymers agrees with SCFT predictions after equilibration, and the incorporation of the mechanophore does not change the overall mechanical response of the material.
We find that in triblocks with higher fractions of the glassy end blocks, more stress is required to activate the mechanophore unit in the midblock than is required in triblocks with lower glassy block fractions.
Additionally, we find that the chain topology is critical, with mechanophores located in tie chains activating at much higher rates than mechanophores located in loop chains. We also find that hooked loop chains behave very similar to tie chains, showing similar stress-activation curves. 
This work suggests that triblock copolymers may offer an attractive platform for controlling mechanochemical activation, and should motivate further experimental work on these systems.
Finally, this work highlights the key role of network topology in determining mechanochemical activation, which may offer new opportunities to understand and optimize mechanochemical responses across a wide range of related systems.

\section{Supplemental Material}
See the supplementary material for the additional simulation
models and summarized data.

\section{Data Availability} 
The data that support the findings of this study are available
from the corresponding author upon reasonable request.

\begin{acknowledgments}
This work was supported by a grant from the National Science Foundation (DMR-1846665). This research was supported in part by the University of Pittsburgh Center for Research Computing through the resources provided.
\end{acknowledgments}

\section{References}
\bibliography{bibliography}

\end{document}


\title{Supplemental Information: Computational Study of Mechanochemical Activation in Nanostructured Triblock Copolymers}
\author{Zijian Huo}
\affiliation{Department of Chemistry, University of Pittsburgh, 219 Parkman Ave., Pittsburgh, PA, USA}
\author{Stephen J. Skala}
\affiliation{Materials Science and Engineering, Grainger College of Engineering, University of Illinois, Urbana-Champaign, IL 61801 }
\author{Lavinia Falck}
\affiliation{Department of Chemistry, University of Pittsburgh, 219 Parkman Ave., Pittsburgh, PA, USA}
\author{Jennifer E. Laaser}
\affiliation{Department of Chemistry, University of Pittsburgh, 219 Parkman Ave., Pittsburgh, PA, USA}
\author{Antonia Statt}
\email[email:]{statt@illinois.edu}
\affiliation{Materials Science and Engineering, Grainger College of Engineering, University of Illinois, Urbana-Champaign, IL 61801 }

\maketitle

\section{Model}
\subsection{General Chain parameters and set-up}

The experimental triblock ABA copolymer was represented by the classical Kremer-Grest coarse-grained model to carry out the molecular dynamic simulations \cite{Grest1990,Grest1986, Kremer1990}. In this model, the polymers are treated as linear chains formed by beads connected by springs. A group of 6-8 chemical monomer units constitutes each bead (see Table~\ref{tab:parametrization}), which is on the length scale of the Kuhn segment in the real copolymer.

Coarse-graining of the model was done using a top-down approach. 
First, the size of each repeating unit needed to be determined. The size ratios of the PMMA and PnBA beads were estimated using the ratio of the Kuhn segment lengths, $l_\text{K}$, which is between 1.12 and 1.28 according to literature values summarized in Table~\ref{tab:parametrization}. Additionally, the Kuhn segment molecular weight ratio was between 1.33 and 1.61, and the melt density ratio was between 0.86 and 0.95. From those values, $m_{B}=1.6m_{A}$ and $\sigma_{B}=1.2\sigma_{A}$, were chosen to give consistent relative masses, relative sizes, and relative melt densities of the two polymers.

For non-bonded interactions, standard Lennard-Jones (LJ) potentials~\cite{Jones1924}
\begin{equation}
U_{\text{pair},ij}(r)=\begin{cases}
          4\epsilon_{ij}\left[\left(\frac{\sigma_{ij}}{r}\right)^{12}-\left(\frac{\sigma_{ij}}{r}\right)^{6}\right] &\text{if} \,\,\, r < r_\text{cut} \\
          0  &\text{if} \,\,\, r \geq r_\text{cut} 
     \end{cases},
\end{equation}
were used, where the relative interaction energies $\epsilon_{AA},\epsilon_{BB}$ determine the physical properties of each polymer, and the cross-interaction $\epsilon_{AB}$ determines the phase separation. 
The bonded interactions were given by the standard finite extensible nonlinear elastic (FENE) potential~\cite{Grest1990,Grest1986, Kremer1990} 
\begin{align}
U_{\text{bond},ij}(r)=-\frac{K_{ij}}{2}R^{2}_{0,ij}\ln\left[1-\left(\frac{r}{R_{0,ij}}\right)^{2}\right] \nonumber \\ +4\epsilon_{ij}\left[\left(\frac{\sigma_{ij}}{r}\right)^{12}-\left(\frac{\sigma_{ij}}{r}\right)^{6}\right] +\epsilon_{ij} \quad .
\end{align}

The energy interaction parameters $\epsilon_{ii}$ were estimated using the experimental glass transition temperatures of PMMA and PnBA, as stated in Table~\ref{tab:parametrization}. The ratio of glass transition temperatures for these two polymers is approximately 0.6, leading to $\epsilon_{BB}=0.6\epsilon_{AA}$.

\begin{table}\centering \caption{Values used for model parametrization}
\begin{threeparttable}
\begin{tabular}{l l l l l} 
 	\hline
   & PMMA &\phantom{X} & PnBA \\
 	\hline\\[-2ex] 
 	$\text{T}_\text{g}$ $[\si{\kelvin}]$ & 367 -- 395\tnote{$*$} & & 229\tnote{$*$,$\dagger$}   \\
Melt density  $ [\si{\gram / \milli \liter}]$ & 1.13\tnote{$\ddagger$} -- 1.2\tnote{$*$} & &	1.03-1.07\tnote{$*$} \\
 	Kuhn segment length $[\si\angstrom]$ & 15.3\tnote{$\mathsection$,$\ddagger$} & & 17.1 -- 19.6\tnote{$i$}  \\
 	Characteristic ratio & 7.3-- 8.2\tnote{$\|$}  & & 9.0\tnote{$j$} - 10.4\tnote{$i$}  \\
 	Kuhn segment weight $[\si\gram/\text{mol}]$ & 598\tnote{$\ddagger$} -- 666\tnote{$+$} & & 844 -- 960\tnote{$i$}  \\  Monomers per Kuhn segment & 5.97\tnote{$\ddagger$}& & 6.6 -- 7.5  \\
 	\hline
 \end{tabular}
\begin{tablenotes}\footnotesize
\item[$*$]Ref.\citenst{wypych2016handbook}
\item[$\dagger$]Ref.\citenst{Zabet2017}
\item[$\ddagger$]Ref.\citenst{fetters2007chain}
\item[$i$]Ref.\citenst{Ahmad2001}
\item[$\mathsection$]Refs.\citenst{mark2007physical,Kuhlman2006}
\item[$\|$]Ref. \citenst{Lu2017}
\item[$+$]Ref.\citenst{Guo2006} 
\item[$j$]Ref.\citenst{Aharoni1985}
\end{tablenotes}
\end{threeparttable}
\label{tab:parametrization}
\end{table}

The cross interaction between $A$ and $B$ beads was determined by the Lorentz~\cite{Lorentz1881} and Fender-Halsey mixing rules~\cite{Fender1962}, resulting in $\sigma_{AB} = \frac{1}{2}(\sigma_{AA}+\sigma_{BB}) = 1.1\sigma$ and $\epsilon_{AB}=2\cdot\epsilon_{AA}\epsilon_{BB}/(\epsilon_{AA}+\epsilon_{BB})=0.75\epsilon$. This value for $\epsilon_{AB}$ did result in phase separation for the systems of interest. In this work, the $\chi N$ parameter was not determined. 

All bonded interactions between PMMA and PnBA beads were given by the FENE potential with the same parameters for $\sigma_{i}$ and $\epsilon_{i}$ as for the pair interactions, leading to $R_{0,ij}=1.5\sigma_{ij}$ and a standard spring constant of $K=30\,\epsilon/\sigma^2$ for all bonds to avoid unphysical bond crossing.

\subsection{Double well potential}

\begin{figure}[!ht]
\begin{center}
\includegraphics[width=7.5cm]{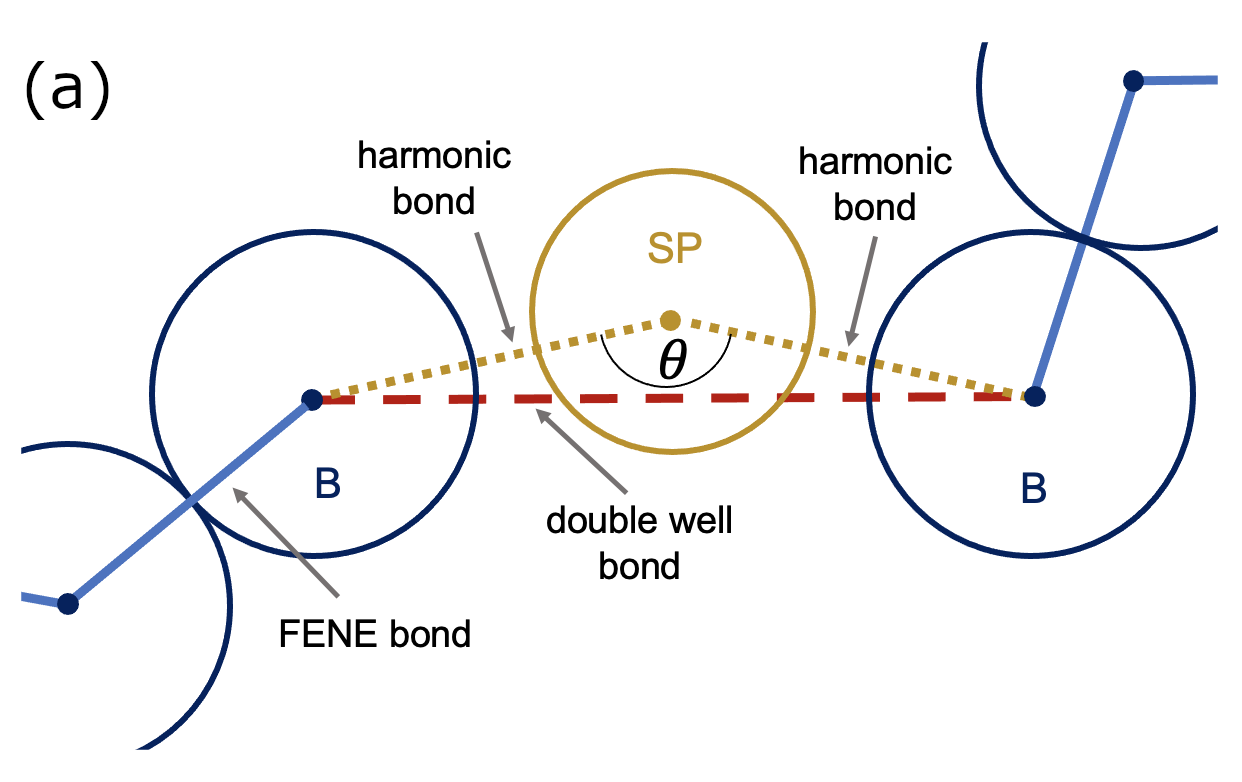}\\[-1em]
\includegraphics[width=7.5cm]{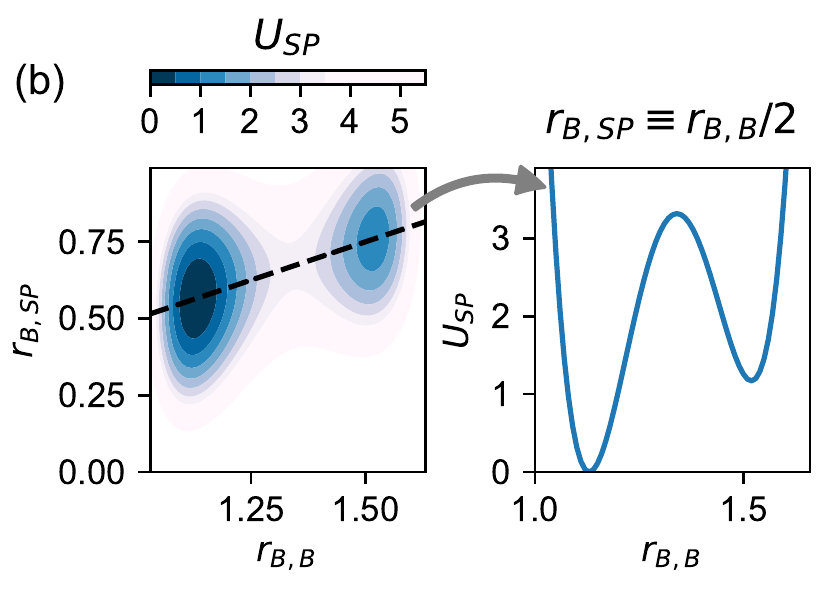}
\end{center}
    \caption{(a) Sketch of the bonds and angles for the mechanophore bead. (b) Left: Energy $U_\text{SP}$ of the bonds on the mechanophore as function of distance $r_{B,B}$ of the two adjacent $B$ beads and distance $r_{B,SP}$  of one $B$ bead to the mechanophore bead $SP$. Right: Energy $U_\text{SP}$ for configurations where the mechanophore is exactly in the middle between the two $B$ beads, as indicated by the dashed line in the left figure. The energies are given by the addition of the double well and harmonic potentials.}
    \label{fig:sketch_mechanophore}
\end{figure}

The mechanophore (SP) was modeled using a coarse-grained bead located in the center of the B block with a radius $\sigma_{SP} = 0.88\,\sigma$. A sketch of the configuration is shown in Fig.\ref{fig:sketch_mechanophore}(a). This size was chosen to prevent bond crossing events. The angle between the SP and adjacent B beads was constrained with a harmonic angle potential,  $U(\theta) = \frac{k_{a}}{2}(\theta - \theta_{0})^{2}$ with a harmonic angle potential constant $k_{a} = 50.0\,\epsilon$ and a rest angle $\theta_{0} = \pi$, to ensure the beads are connected in an approximately straight line.

To model the activation of a mechanophore, a double well potential
\begin{equation*}
U_{DW}(r) = \frac{U_{max}}{b^{4}}\left[\left(r-\frac{a}{2}\right)^2-b^{2}\right]^{2}\quad,
\end{equation*}
between the two adjacent $B$ beads was implemented. In this potential,
$U_{max}$ is the energy required for activation, $\frac{a}{2} \pm b$ are the location of the two minima, and $\frac{a}{2}$ gives the location of the energy barrier. 

The interaction between the SP bead and the other beads (A, B, SP beads) were calculated using the Lennard-Jones potential with interaction strength $\epsilon = 1.0\,\epsilon$ and cutoff radius $r_{c,ij} = 2^{1/6}\,\sigma_{ij}$, with the radius, $r_{ij}$, determined by the same mixing rules as before. The bonded interactions between the SP bead and the adjacent B beads were set using a harmonic potential $U_{bond}(r) = \frac{k_{b}}{2}(r - b_{0})^{2}$ with spring constant $k_{b} = 15\,\epsilon/\sigma^2$ and equilibrium distance $b_{0} =  \frac{a}{4} -  \frac{b}{2}$ chosen to match half of the location of the first minimum in the double well potential, such that both harmonic bonds kept the SP bead in the middle in equilibrium, as shown in Fig.~\ref{fig:sketch_mechanophore}(a). The double well and harmonic bond potentials together are shown in Fig.\ref{fig:sketch_mechanophore}(b). Because the harmonic bonds are stretched away from their equilibrium position when the SP bead is in the second double well minimum, they effectively ``tilt'' the energy of the SP bead, favoring the first minium by roughly $1\,\epsilon/k_\text{B}$.  


\begin{figure}[!h]
    \centering
    \includegraphics{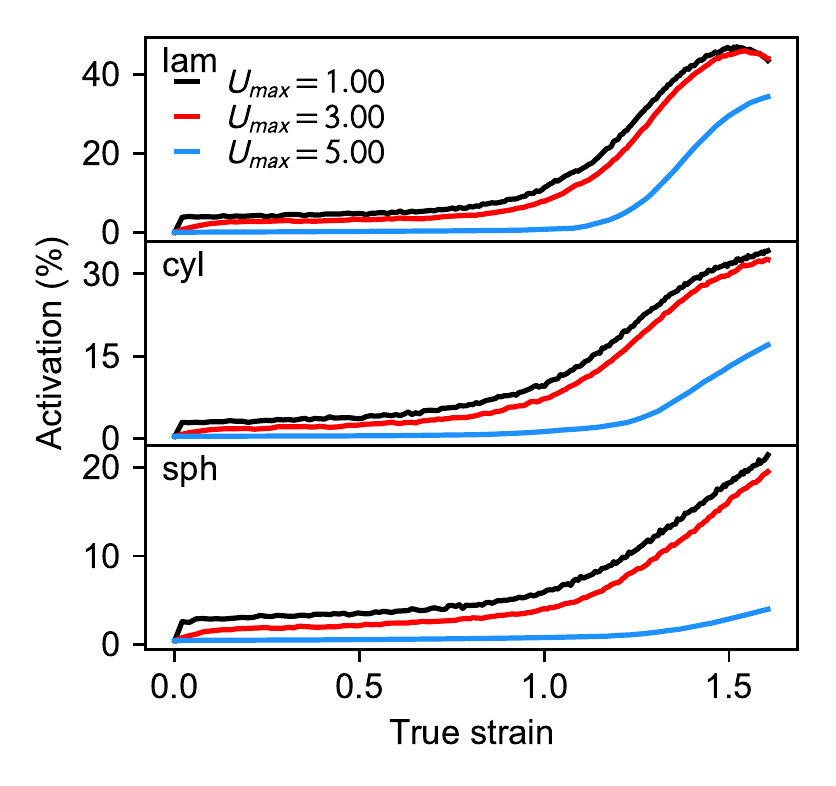}
    \caption{The activation of the mechanophore with various values of $U_{max}$ in units of $\epsilon/k_\text{B}$. While the absolute activation changes, the trends between morphologies does not.}
    \label{fig:SP_act_v}
\end{figure}
    As the potential height, $U_{max}$, chosen for the SP double well potential was somewhat arbitrary, it was varied to determine how it affects the activation of mechanophores. As shown in Fig. \ref{fig:SP_act_v}, the trends of the activation for all morphologies were approximately the same for different values of $U_{max}$; however, the absolute activation varied significantly, where the mechanophores with lower barrier heights activated more strongly than the ones with higher barrier heights.
    Additionally, we note that changing the potential barrier had a stronger impact on the spherical morphology than on the spherical or lamellar morphologyies, although the overall trends in activation again remained unchanged.
    
\section{Methods}
\subsection{Morphology Assembly and Equilibration}

The desired morphologies were obtained by varying the volume fraction, $f$, of A beads in the triblock copolymer from 0.16 to 0.50 as shown in Table~\ref{tab:systems}. The chains were randomly placed in a cuboid box of volume $V = L_{x} \times L_{y} \times L_{z}$ with standard periodic boundary conditions at a number density of $0.445\,\sigma^{-3}$.

\begin{table}[!h]
\begin{center}
\caption{Polymer Samples, volume fractions $f_A$, box sizes, total number of chains $N_{chains}$.\label{tab:systems}}
\begin{tabular}{cccccc}
\hline
& Sample ID & Morphology & $f_A$  & Box &  $N_\text{chains}$\\ \hline
& $A_{8}B_{50}A_{8}$ & spherical & 0.16 & $72 \sigma \times 72 \sigma  \times 72 \sigma $ &  3813 \\
& $A_{15}B_{50}A_{15}$ & cylindrical & 0.26 & $80 \sigma  \times 69 \sigma  \times 40 \sigma $ & 2003  \\
& $A_{43}B_{50}A_{43}$ & lamellae & 0.50 & $92 \sigma  \times 92 \sigma \times 46 \sigma $ & 2373 \\
& $A_{43}B_{50}A_{43}$ & random & 0.50 & $80  \sigma  \times 80 \sigma  \times 40 \sigma $ & 1262 \\
\hline
\end{tabular}
\end{center}
\end{table}

An external field was applied on the system to facilitate faster equilibration from the disordered state to the desired morphologies, inspired by Padmanabha et. al.'s approach in Ref.[ \citenum{Padmanabhan2016}]. The field attracted the A beads and had no interactions with the B beads. This field interaction was modeled using discrete grid points on a square lattice with the lattice constant of $1.0\sigma$, each interacting with the A beads via a Gaussian potential:
\begin{equation}
U_{\text{gauss}}(r)=\begin{cases}
          \epsilon_{XA}\,\text{exp}\left[-\frac{1}{2}(\frac{r}{\sigma_{XA}})^{2}\right] &\text{if} \,\,\, r < r_\text{cut} \\
          0  &\text{if} \,\,\, r \geq r_\text{cut} 
     \end{cases},
\end{equation}
Those parameters were chosen to be $\epsilon_{XA} = -0.1\epsilon$, $\sigma_{XA} = 0.4\sigma$, and $r_{cut} = 2\sigma$, resulting in a net attraction of A beads towards the desired discretized field. The fields were modeled according to the expected morpolgies with domain spacings measured from direct unbiased simulations. The box sizes (see Table \ref{tab:systems}) were chosen to be commensurate with the domain spacing. The final morphologies were not sensitive to the exact parameters chosen for the discretized field, as long as the field beads attracted the A beads.

The morphologies in the external field were then obtained using dissipative particle dynamics (DPD) with repulsive parameters $a_{AA} = a_{BB} = 30.0$ and $a_{AB} = 60.0$, and standard parameters otherwise.~\cite{Phillips2011,Groot1997} 
The bonded interactions were described by a harmonic potential, $U_{bond}(r) = \frac{k_{0}}{2}(r - b_{0})^{2}$, with the equilibrium lengths being equal to the equilibrium bond lengths from the FENE and WCA potential and the spring constant being an arbitrary value of $k_{0} = 100.0\,\epsilon/\sigma^2$. 

The equations of motion were integrated using the velocity-Verlet algorithm with a timestep of $\Delta t = 0.04\,\tau$. The DPD equilibration simulations were performed in the constant volume and constant temperature (NVT) ensemble for $1.2 \cdot 10^{4}\tau$ at a temperature of $T= 1\,\epsilon/k_{B}$. After the desired morphology was achieved, the external field was removed and the DPD repulsive parameters were increased to $a_{AA} = a_{BB} = 100.0$ and $a_{AB} = 500.0$, and the harmonic bond spring constant was increased to $k_{0} = 500$. The timestep was decreased to $\Delta t = 0.001\,\tau$ and a FIRE energy minimization~\cite{FIRE} was done for $500\tau$ to remove any remaining overlaps.

After the desired microstructure was achieved and overlaps removed, the soft DPD potential was replaced with the Lennard-Jones potential and the harmonic bond potential was replaced with the FENE and WCA potentials as previously described. The timestep was also adjusted to $\Delta t = 0.005\,\tau$. The sample was then relaxed in the constant volume and constant temperature (NVT) ensemble using a Langevin thermostat for $2.5 \cdot 10^{3}\tau$ to further equilbrate after changing the potential. 

\begin{figure}[!ht]
    \centering
    \includegraphics{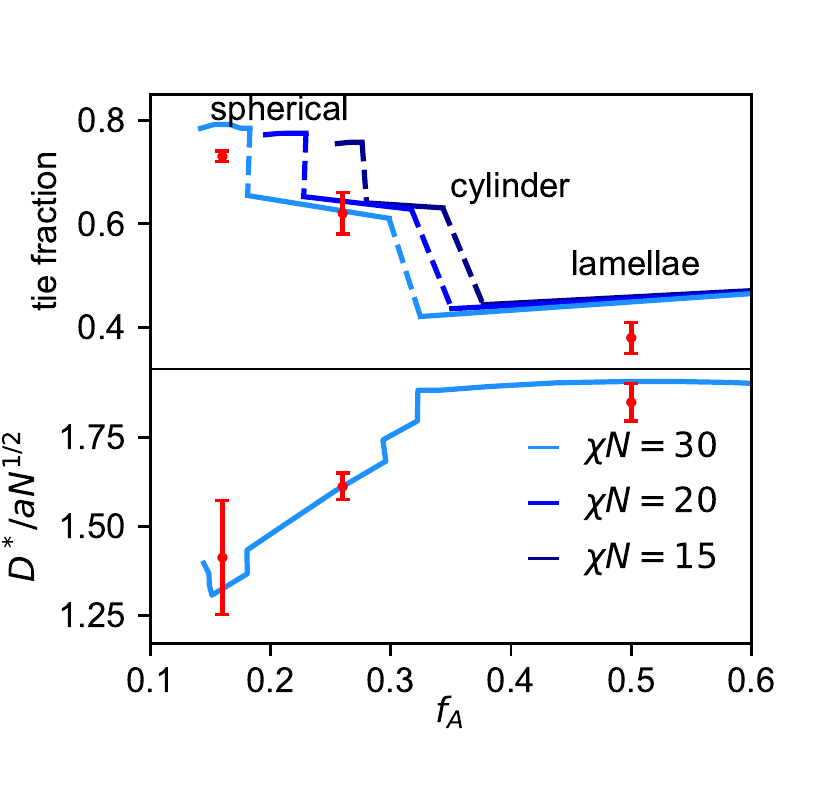}
    \caption{Top: Tie fractions and Bottom: Normalized domain spacings as measured from simulations (red points) compared to SCFT results (blue lines) from Ref.~[\!\citenum{Matsen1999}].}
    \label{fig:equi_loop_tie_domain_spacing}
\end{figure}  
The system was then cooled to the desired temperature of $T = 0.35\,\epsilon/k_{B}$, which is below the glass transition temperature ($T_{g}$) of the A blocks \cite{Zabet2017}, under the constant pressure and constant temperature (NPT) ensemble using a MTK barostat-thermostat with coupling constants $\tau = 1.0$ and $\tau_{p} = 1.2$ at constant pressure $p = 0$. The system was cooled at a cooling rate of $\dot{\Gamma}$ of $1.44 \cdot 10^{-4}\epsilon/k_{B}\tau$  for $4.5 \cdot 10^{3}\tau$. The system was then equilibrated at $T = 0.35\,\epsilon/k_{B}$ for another $500\tau$.

The mechanophore units were then introduced into the system as described in the model section and the system was deformed as described in the main text.

In Fig.\ref{fig:equi_loop_tie_domain_spacing}, the tie fractions and  domain spacings as measured from simulations (red points) and SCFT~\cite{Matsen1999} (blue lines) are shown. In the lamellar sample, $D^*$ is the spacing between lamllae; in the cylindrical sample, the spacing between cylinders is given by $(4/3)^{1/2}D^*$; and in the spherical morphology, the distance between spheres is $(3/2)^{1/2}D^*$. For $a$, the monomer bead size, we used $1.2\sigma$, i.e., the bigger rubbery bead of the majority phase. 
Because the exact value of $\chi N$ was not determined in this study, and the beads for each block were not identical, small discrepancies are expected. Because the middle block was kept fixed in size with $N_B=50$, resulting in different total chain lengths $N$, each morphology has a different $\chi N$ value, increasing from the spherical to the lamellar morphology. Overall, the good agreement between SCFT predictions and the simulation results indicated that the simulation method for assembling the morphologies generated equilibrated samples with the expected chain conformations.

\subsection{Primitive Path Algorithm and Chain Identification}

We used the primitive path analysis algorithm~\cite{Karayiannis2009,Shanbhag2007,Kroger2005,Foteinopoulou2006} to obtain the contour lengths and kinks on each rubbery $B$ mid-block. To determine the primitive path, we applied the geometric method to find the shortest path on the $B$ beads only. In short, the algorithm reduces the polymer contour (as given by the bond vectors) while keeping all entanglements and intersections intact. This is achieved by a series of geometric operations on each segment (as given by the bond vectors), removing the intersection-free segments, and reducing other segments to their convex hulls given by intersection points with all other chains. This series of geometric operations is repeated until the path cannot be simplified any further. We refer to Refs.~[\citenum{Kroger2005,Shanbhag2007}] for a detailed discussion of the algorithm. The total number of beads or segments of each chain is not conserved; instead, the remaining points at the end of the algorithm indicate kink or entanglement positions. From the primitive path of the rubbery B mid-block, it is then straightforward to calculate the number of kinks $k = N_\text{segments}-1$, as well as the length of the primitive path $L_{c,B} = \sum L_\text{segments}$. We note that we found energy minimization or annealing schemes for determining the primitive path~\cite{Shanbhag2007,Sukumaran2005} less suitable, because the algorithm stability and its adjustable parameters were dependent on the current elongation and morphology, whereas the geometric method was stable and could be applied to all morphologies over the entire deformation trajectory with no adjustable parameters.

Different chain types can be then classified once the chains are reduced to their primitive path. Tie chains are chains connecting two neighboring glassy clusters or domains in the morphology. To identify tie chains, we determined the clusters in the morphology using the DBSCAN algorithm~\cite{dbscan1996} on the glassy $A$ beads with a nearest-neighbour cutoff of $1.3\sigma_{A}$ and a minimum cluster size of $N_\text{A}+1$, such that isolated dangling chains are not counted as clusters.  Because our samples were well separated and ordered, the DBSCAN algorithm was able to clearly identify all clusters in all morphologies. The cluster identifcations did not depend on the neighbour cutoff, when chosen within a reasonable range. Therefore, we picked the first minimum in the pair correlation function of the $A$ beads, $\approx 1.3\sigma_A$, as the cutoff. 
Tie chains were then identified simply as those chains connecting two separate glassy clusters, and dangling chains were identified as the chains with one end being part of a cluster while the other one was not. Additionally, we identified loop chains as chains where both ends are in the same cluster. 

To determine if a given loop chain was hooked to another loop chain or not, we calculated the writhe~\cite{CIMASONI2001,Young2021}, or twist number, for all loop chains with all other loop chains. The writhe $Wr_{\alpha,\beta}$ between two rubbery segments is given by a double line integral 
\begin{align}
    Wr_{\alpha,\beta} &= \frac{1}{4\pi} \int_{C_\alpha}  \int_{C_\beta} d\bf{r_1} \times d\bf{r_2} \cdot \frac{\bf{r}_1-\bf{r}_2}{\abs{\bf{r}_1-\bf{r}_2}^3} \\
    &\approx \frac{1}{4\pi}  \sum_i \sum_j \frac{\bf{r_i}-\bf{r_j}}{r^3_{ij}} \bf{r_i} \times \bf{r_j}\quad,
\end{align}
where we approximated the line integrals over the continuous polymer curves $C_\alpha$ and $C_\alpha$ by the discrete positions of the primitive path. If $\abs{Wr_{\alpha,\beta}}>0$ and the chain ends of $\alpha$ and $\beta$ are in different glassy clusters, we considered the rubbery mid-blocks $\alpha$ and $\beta$ to be hooked. 

To summarize, we analyzed every snapshot in our simulations by determining the glassy clusters and the primitve path of the rubbery blocks. With this approach, we identified
\begin{description}
\item[tie chains:] chains connecting two neighboring glassy domains,
\item[dangling chains:] chains with one A-block end which is not part of a glassy cluster,
\item[loop chains:] chains whose ends connect to the same cluster, and which are not hooked to another loop chain, and
\item[hooked chains:]  chains whose ends connect to the same cluster, and which are hooked to another loop chain from a different cluster.
\end{description}
Obviously there are many other, more complicated topological states chains could be in, e.g. a loop chain can be entangled or hooked to a tie chain, etc., but we have limited our analysis to the ``hooked'' chains as being the most obvious one and leave further, more detailed analysis to future work.

\section{Additional Results}
\subsection{Slow Strain Rate}

\begin{figure}[!h]
    \centering
    \includegraphics[width=8.5cm]{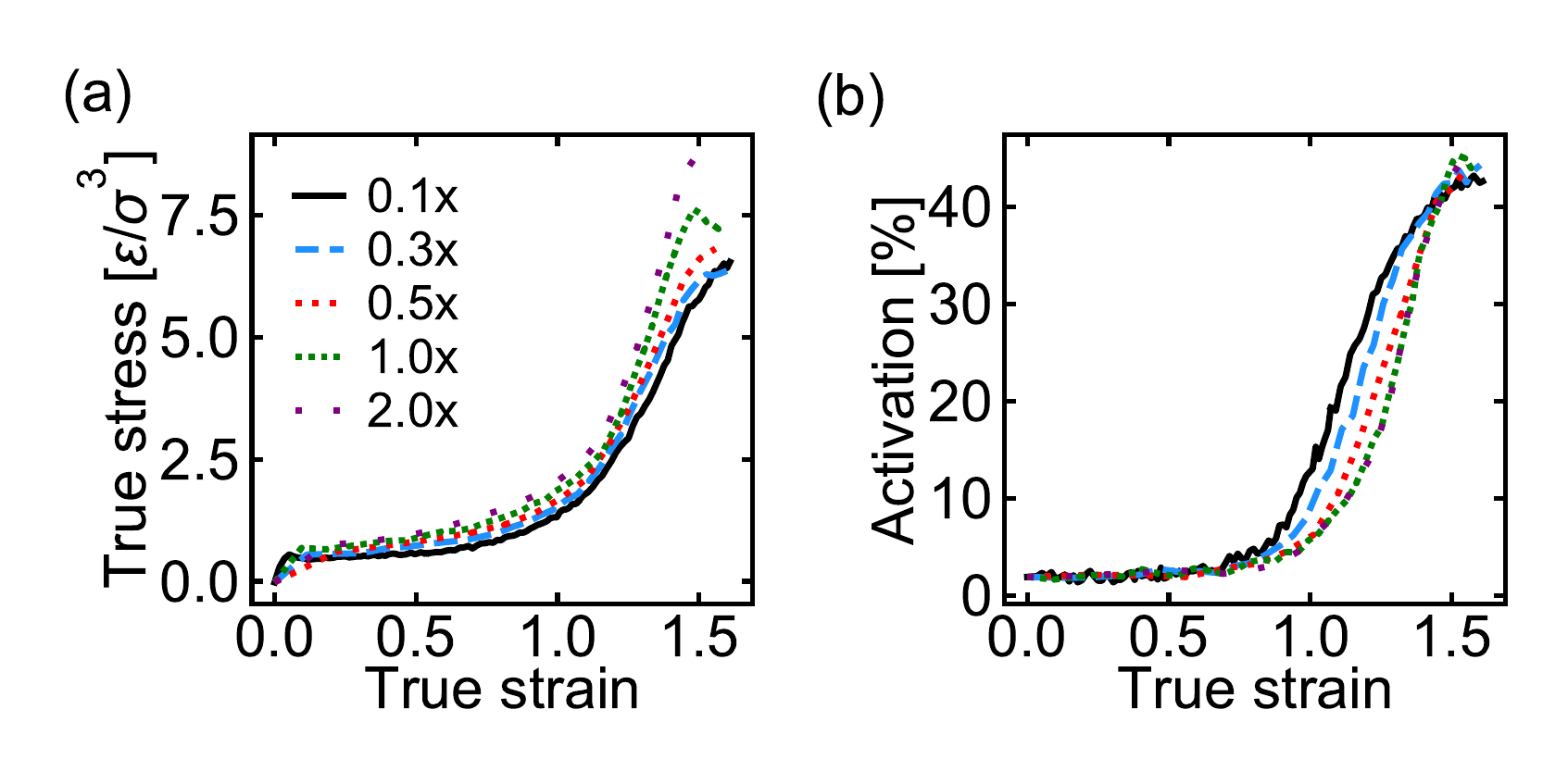}
    \caption{Stress-strain and activation-strain behaviors of the deformation of a lamellar sample by varying the strain rate. The 1x curve represents the strain rate used for the deformation reported in the main text; this curve was averaged over ten independent simulation runs, while the other curves representing the relative strain rates to the chosen one were each calculated from a single simulation run.}
    \label{fig:slow_strain_rate}
\end{figure}

As the simulated strain rate ($10^7$-$10^8$ $s^{-1}$) was orders of magnitude higher than experimentally possible rates ($10^{-3}$-$10^{-1}$ $s^{-1}$), simulations were tested at five different strain rates for a lamellar sample to determine how the strain rate affects mechanophore activation. As shown in Fig. \ref{fig:slow_strain_rate}, the lamellar sample exhibited similar glassy polymer stress-strain behaviors at each strain rate studied. At each strain rate, a very short elastic regime followed by a strain hardening was observed. At faster rates, the samples showed higher stresses, as expected,
but slower rates gave higher activation. We speculate that slower strain rates give more time for the polymer chains to stretch under deformation, which allows more mechanophores to activate. 


\subsection{Ordered vs disordered structure}

\begin{figure}[!h]
    \centering
    \includegraphics[width=8.5cm]{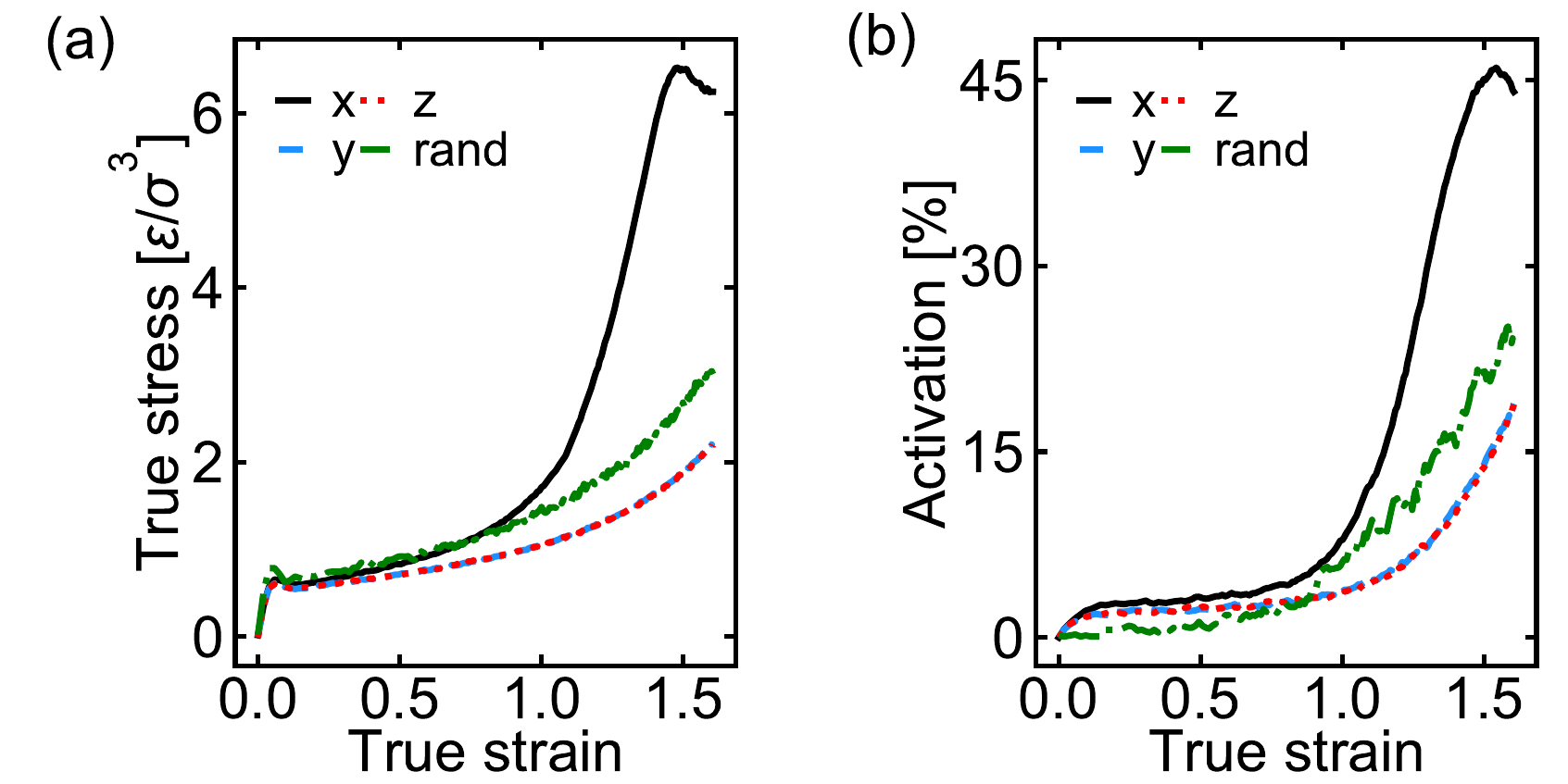}
    \caption{(a) Stress strain and (b) activation curves of lamellar morphology deformed in different directions, compared to response for a sample with the same composition but random morphology.  The curves for the lamellar sample were averaged over ten independent simulation runs, while the curve for the random sample was calculated from a single simulation run. }
    \label{fig:stress_strain_activation_with_random}
\end{figure}

To probe the influence of the ordered structure in activation, a random sample was generated by increasing $\epsilon_{AB}$ from 0.75 to 1 and skipping DPD steps to prevent large scale phase separation. All other parameters were identical to those used to generate the lamellar morphology as described earlier. Both ordered and disordered samples were deformed in the same way, and resulting stress strain curves are displayed in Fig.\ref{fig:stress_strain_activation_with_random}.

The random sample showed a similar stress strain behavior to the lamellar sample, with stresses between the values measured when the lamellar sample was deformed perpendicular ($x$) and parallel ($y$, $z$) to its layers. The random sample also showed lower activation. This suggests that the well-ordered morphology can increase the activation of a mechanophore \textit{if} the polymer chains align with the deformation direction. 

\subsection{Chain Alignment}

\begin{figure*}[!t]
    \centering
    \includegraphics{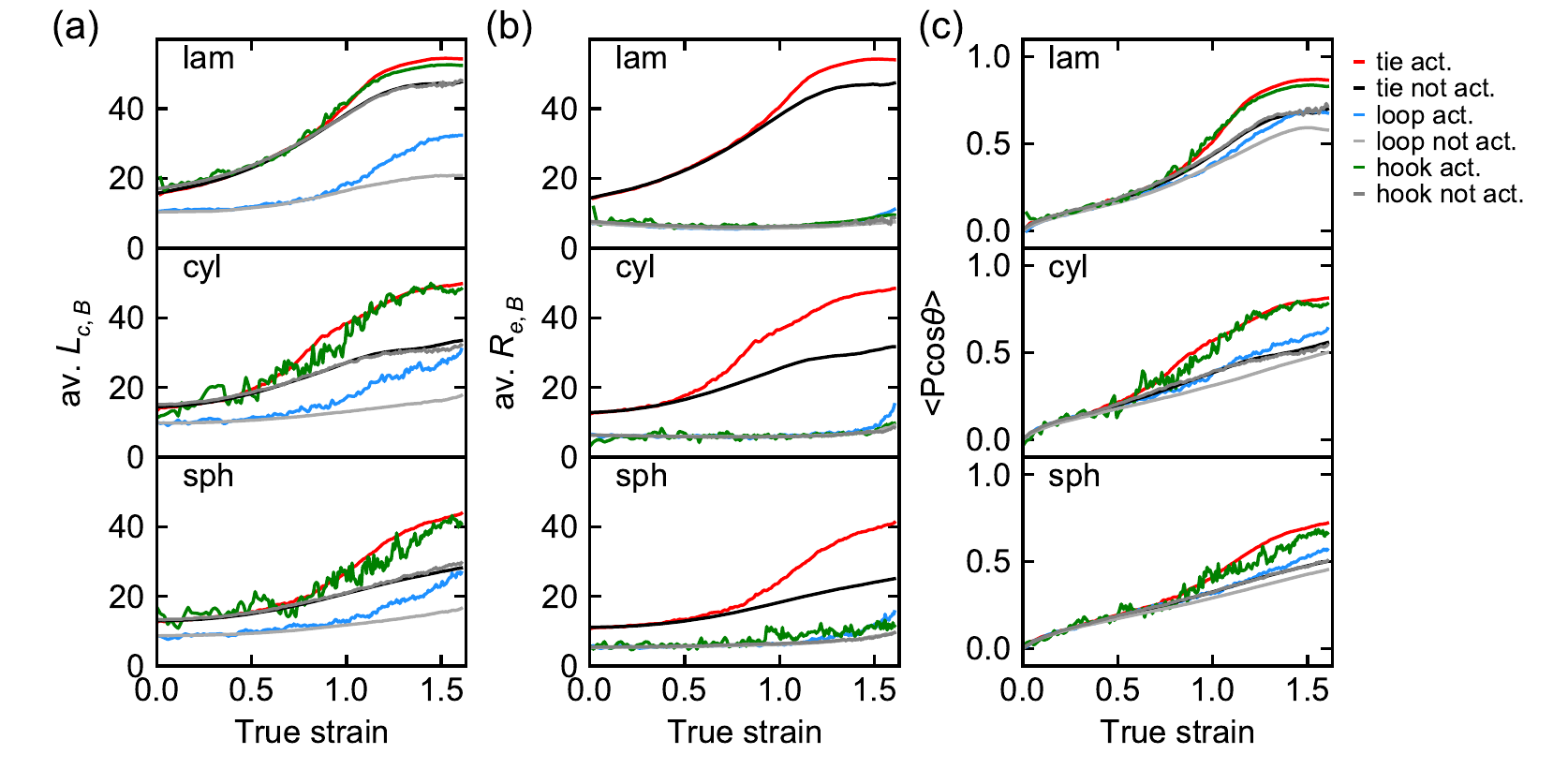}
    \caption{ The change of the (a) average primitive path contour length $L_{c,B}$ and (b) average end-to-end distance $R_{e,B}$ of B midblocks for the different polymer chain types. (c) BB bond orientations parameter, $\langle P\cos{\theta}\rangle $. The legend to the right is for all panels.} 
    \label{fig:ppa_over_time}
\end{figure*}

In Fig.~\ref{fig:ppa_over_time}, both the average primitive path length and end-to-end distance for the B midblock are plotted to quantify the evolution of the chain conformations during deformation. 
For low strain values, the curves for the activated and the unactivated chains overlap, and there was no significant difference in either the end-to-end distance or the primitive path length. As expected, for low strains (and in equilibrium), the tie chains had the highest $L_{c,B}$ and $R_{e,B}$ in all morphologies.  

From the average values of $L_{c,B}$ and $R_{e,B}$ as function of strain (Fig.~\ref{fig:ppa_over_time}), it is immediately apparent that activated tie chains had the longest extension of all chain types, especially at high strain. Hooked chains had a low $R_{e,B}$ overall, but their primitive path length increased significantly with strain, with overall values comparable to the tie chains. This illustrates that hooked chains behaved very similarly to tie chains. 
The loop chains had low average $R_{e,B}$ as well, which only increased slightly at high strain. Their average primitive path length $L_{c,B}$ did, however, increase somewhat more substantially with strain, especially for activated chains.
   
Overall, the trends in all morphologies are consistent, with the lamellar sample showing having the highest absolute chain extensions at the end of the deformation, followed closely by the cylindrical morphology, with the spherical morphology showed the lowest chain extensions. 
It is possible this is partly due to the fact that it is the easiest to pull out chains from their glassy domains in the spherical morphology because they have the shortest glassy A blocks, however a deeper analysis of dangling chain ends would be necessary.


We also compared the value of $\langle P\cos{\theta}\rangle $, the second order Legendre polynomial, for the bonds in the rubbery region to understand the role of chain alignment in activation under deformation.\cite{1997} The order parameter$\langle P\cos{\theta}\rangle $,describing the average bond orientation in the system, was calculated using
\begin{align}
    P\cos(\theta) = \frac{1}{2}(3\cos^2{\theta} -1)\quad,
\end{align}
where $\cos(\theta)$ is the angle between bonds and the deformation direction. If the bonds are perfectly aligned with the deformation direction, then $\langle P\cos{\theta}\rangle \equiv 1$. 

From Fig.~\ref{fig:ppa_over_time}, we observed that the activated tie and hooked chains are were more aligned with the deformation direction than the loop chains. 
In other words, if the polymer chains are more aligned with the deformation direction, then they are more likely to get activated, as expected. 
The overall increase in alignment  is larger in the lamellar morphology compared to the spherical and cylindrical morphologies, which results in the higher activation observed for the lamellar sample shown in the main text.

\subsection{Kink Analysis}

\begin{figure*}[!t]
    \centering
    \includegraphics{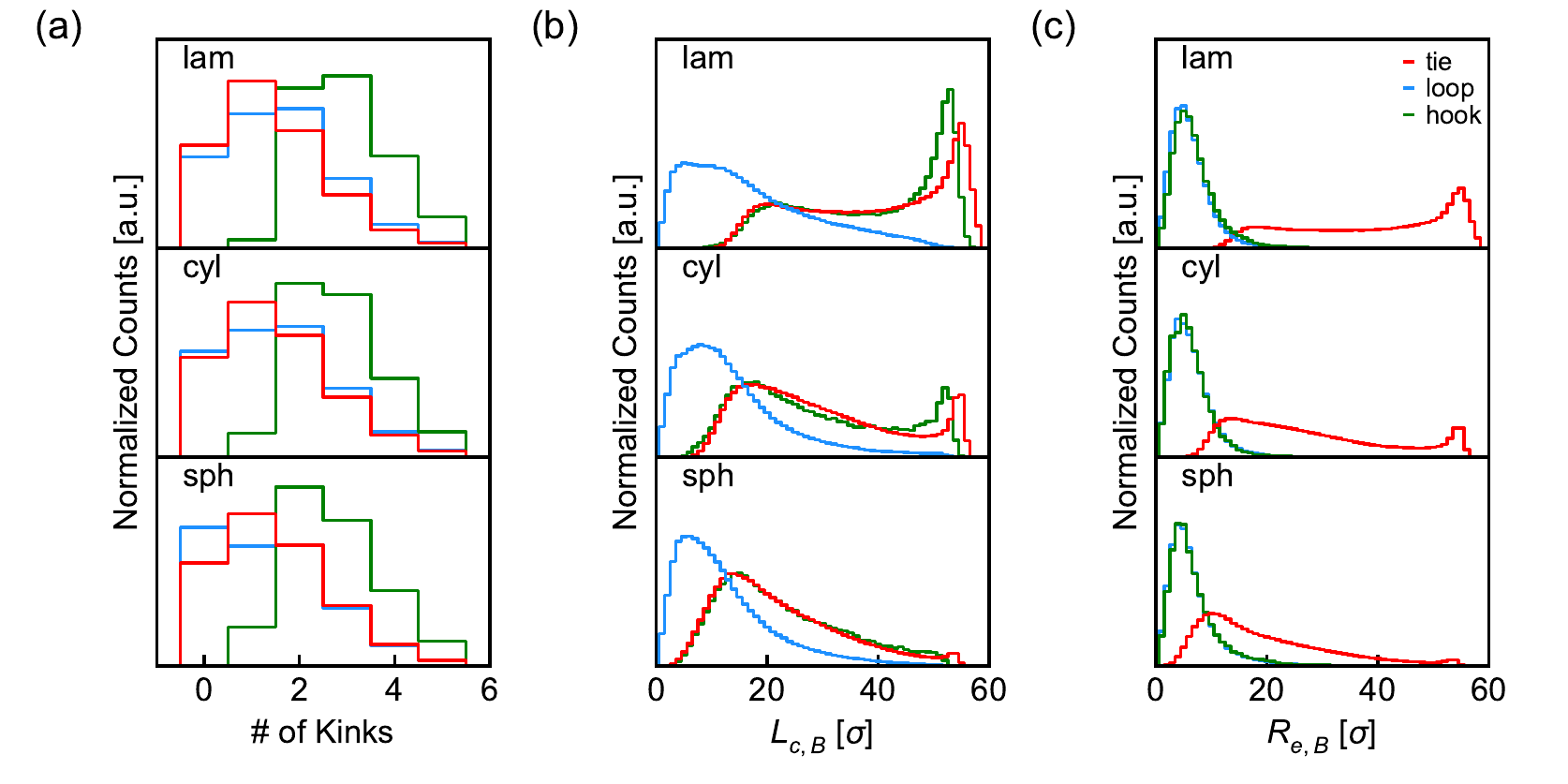}
    \caption{ Histograms of (a)  the number of kinks,  
    (b) the primitive path contour length $L_{c,B}$, and
    (c) the end-to-end distance  $R_{e,B}$ as calculated from primitive path analysis of the $B$ midblocks
    for $f_A=0.5$ lamellar (top panels),  $f_A=0.26$ cylindrical (middle panels), and $f_A=0.1$ spherical (bottom panels) morphologies. For each morphology, the tie (red), loop (blue) and hooked (green) chain histograms are shown.}
    \label{fig:histograms}
\end{figure*}

We observed between zero and 6 kinks on all chains, with averages of $1.62$ for the loop, $1.44$ for tie, and $3.02$ for the hooked chains, as shown in Fig.~\ref{fig:histograms}(a). Because we applied the primitive path analysis to the rubbery block of the polymer chain only, the different morphologies can be directly compared. The main difference between the morphologies was that the loop chains had a slightly lower average number of kinks in the spherical morphology. Note that we have not distinguished between activated/not activated chains in Fig.~\ref{fig:histograms}(a)-(c) and that the histograms contain data for the entire deformation runs.

Fig.~\ref{fig:histograms} also shows the probability distribution of the primitive path length $L_{c,B}$ in panel (b) and end-to-end distance $R_{e,B}$ in panel (c), as well as their ratio $R_{e,B}/L_{c,B}$ in panel (d). Primitive path length histograms show that the loop chains overall had shorter $L_{c,B}$, with a long tail towards higher values. The hooked and tie chain histograms overlap, with some differences in where their peak for the longest extension was, with tie chains extending slightly further than hooked chains on average.  
For the end-to-end distance, plotted in Fig.~\ref{fig:histograms}(c), tie chains displayed higher values, but again with a very broad distribution. The loop and tie chain histograms for $R_{e,B}$ overlap almost perfectly. The hooked chains showed $R_{e,B}$ values similar to those of the loop chains, but $L_{c,B}$ values more similar to those of the tie chains.

\begin{figure}[!h]
    \centering
    \includegraphics{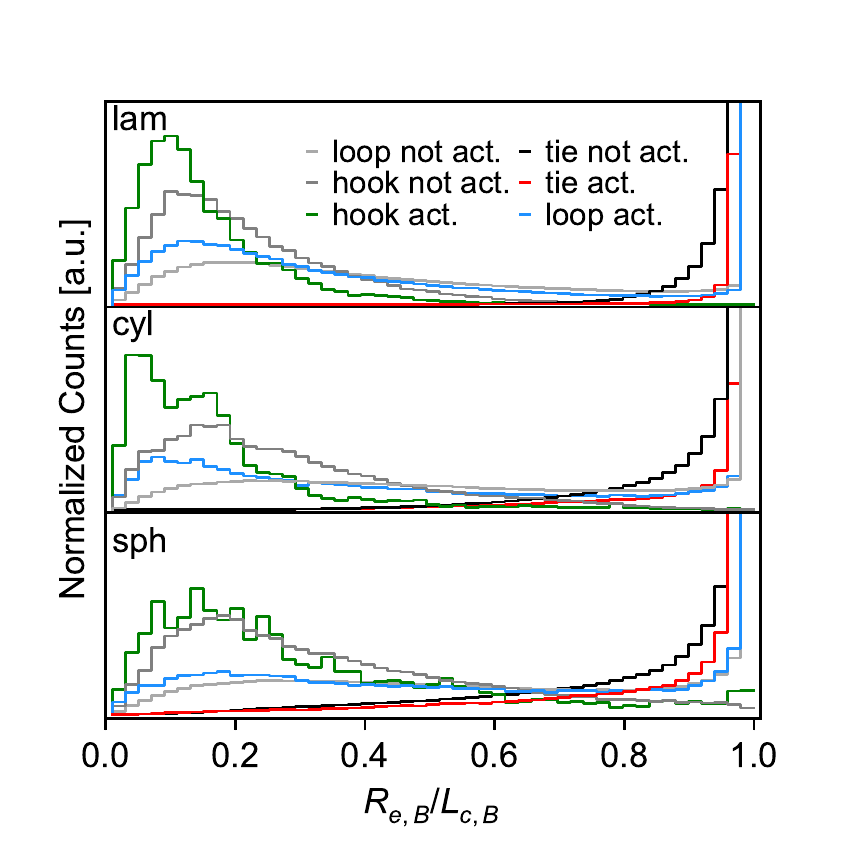}
    \caption{ Histogram of the ratio  $R_{e,B}/L_{c,B}$ as calculated from primitive path analysis of the $B$ midblocks
    for $f_A=0.5$ lamellar (bottom panel),  $f_A=0.26$ cylindrical (middle panel), and $f_A=0.1$ spherical (top panel) morphologies.}
    \label{fig:histograms_ReeLc}
\end{figure}  

In Fig.~\ref{fig:histograms_ReeLc} the ratio of end-to-end distance and primitive path length, $R_{e,B}/L_{c,B}$ is shown. For any chain in any configuration, this ratio can be between one and zero, e.g, entirely straight stretched out configurations with $R_{e,B}\approx L_{c,B}$ have $R_{e,B}/L_{c,B}\approx 1$, while chain configurations with very small $R_{e,B}$ and non-zero $L_{c,B}$ have $R_{e,B}/L_{c,B}\to 0$. For the tie chains (activated and not activated) a prominent peak at $R_{e,B}/L_{c,B}\approx 1 $ is visible, meaning that they adopted mostly ``straight'' primitive path conformations regardless of whether they were activated or not. The loop chains exhibited broader histograms, with a sharp peak at  $R_{e,B}/L_{c,B}\approx 1 $ and a broad peak at $R_{e,B}/L_{c,B}\approx 0.1 \, - \, 0.2$, indicating a substantial number of chains having conformations where the end-to-end distance was significantly shorter than the primitive path length. Geometrically, this means that some loop chains adopt straight primitive path conformations similar to the tie chains, albeit with much shorter total length, as shown in  Fig.~\ref{fig:histograms}(b). Other loop chains adopt more complicated configurations, due to entanglements and kinks, leading to $R_{e,B}<L_{c,B}$. 

The only histograms lacking the peak at $\approx 1$ in Fig.\ref{fig:histograms_ReeLc} were the hooked chains, regardless of activation state. Instead, they showed only the broad peak at $R_{e,B}/L_{c,B}\approx 0.1 \, - \, 0.2$. The peak location shifted somewhat with morphology, from lower values in the lamellar morphology to higher values in the spherical morphology. From a geometrical standpoint, this observation can be rationalized as follows: hooked chains are unable to have perfectly straight primitive paths, as they have at least one entanglement keeping them hooked to another chain from a different glassy domain. For hooked chains, $R_{e,B}<L_{c,B}$. The activated hooked chains showed slightly lower $R_{e,B}/L_{c,B}$ on average, indicating that the hooked chains with larger primitive path length were slightly more likely to be activated.

\subsection{Additional Stress and Activation Plots}

\begin{figure*}[t]
    \centering
    \hspace*{-1.5em}
    \includegraphics[width=5.4cm]{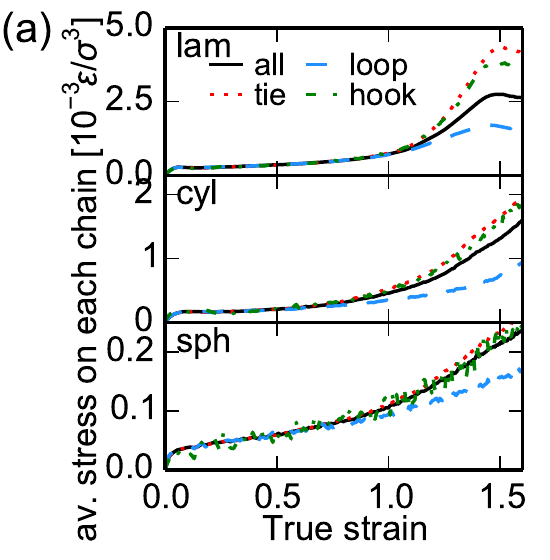}
    \includegraphics[width=5.4cm]{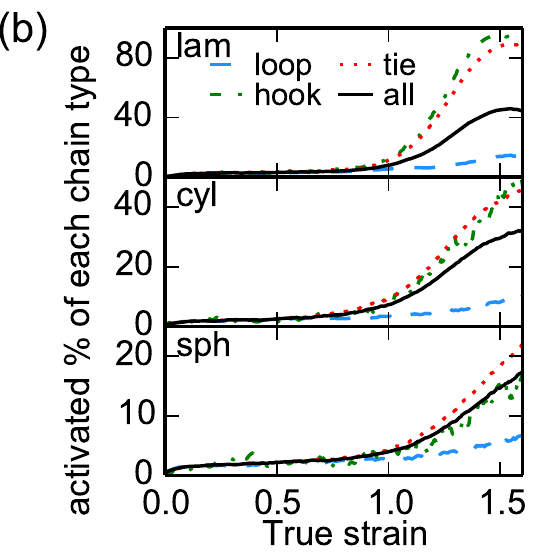}
    \includegraphics[width=7.3cm]{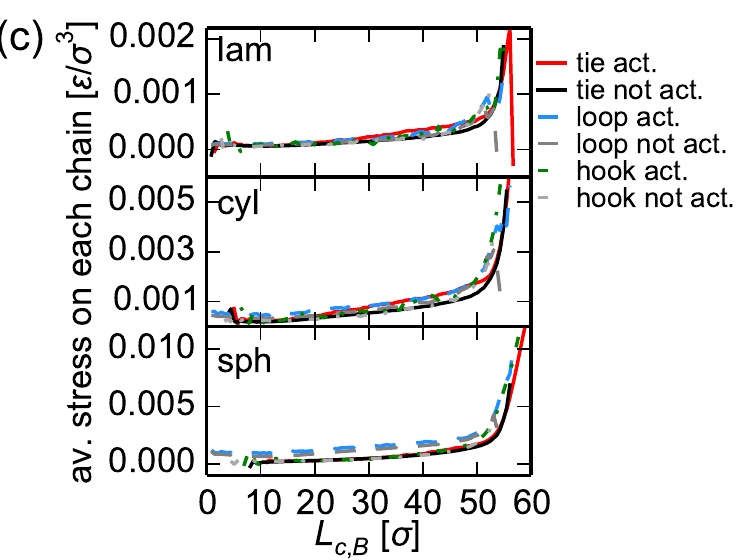}
    \caption{(a) Average stress on each chain as function of chain type, as indicated. Black lines, if multiplied by the total number of chains in the system,  represent the same data as in Fig. 2(a) of the main manuscript.
    (b) Fraction of activated chains per chain type. Black lines represent the same data as in Fig.3(a) of the main manuscript. 
    (c) Average stress on each chain type as function of rubbery primitive path length $L_{c,B}$.}
    \label{fig:stress_different_normalizations}
\end{figure*}

In Fig.~\ref{fig:stress_different_normalizations}(a) the average stress on each chain as function of strain is shown. This data corresponds to Fig.~2 in the main manuscript, but it is normalized differently here, such that the  lines represent average stress values, i.e. divided by number of chains for each chain type. Consequently, one needs to multiply the black lines by the total number of chains in the system to recover the result shown in Fig.~2(a) in the main manuscript. When displayed this way, however, it is clear that the tie and hooked chains had stresses above the average, whereas the loop chains had stresses below the average. The hooked chains behaved very similarly to the tie chains, although their average stress was a little lower than the stress on the tie chains.

Fig.~\ref{fig:stress_different_normalizations}(b) present the same data as Fig.~3 in the main manuscript, but normalized by the number of chains for each type, thus resulting in the activated fraction of each chain type. The black lines correspond to the same data as in Fig.~3(a) in the main manuscript. Again, similarly to Fig.~\ref{fig:stress_different_normalizations}(a), it is clear that loop chains had lower than average activation, whereas tie and loop chains had higher percentages of activation. Interestingly, here, we also observe a trend with morphology: the hooked chains had the highest absolute activation in the lamellar morphology, about the same activation as the tie chains in the cylindrical morphology, and a lower activation in the spherical morphology, close to the overall average. With decreasing glassy fraction in the system, the relative activation of the tie chains thus appears to goes down. This trend is not visible in the average stress on each chain, where the tie and hooked chain essentially behaved identically.

From all of the geometrical properties ($L_{c,B}$, $R_{e,B}$, $R_{e}$, $\#$ kinks) analyzed above, we found that the stress was correlated most strongly with the primitive path length $L_{c,B}$, which is shown in Fig. \ref{fig:stress_different_normalizations}(c). As one might expect, as the primitive path length increased, the stress on each chain increased. The overall shape and behavior of this curve was also universal for all chain types and morphologies: first, an almost linear increase for low values of $L_{c,B}$ and stress was observed, followed by  a sharp increase to high stress values around $L_{c,B}\approx55\sigma$. Those chains with $L_{c,B}\approx 55\sigma$ and above are highly stretched out and therefore experience high values of stress.



In all cases, the average stress on a chain is slightly higher when it is activated, consistent with all other observations made in this work. We also note that a loop chain with the same primitive path length $L_{c,B}$ has a higher stress on it than the other chain types. This is most likely due to the fact that a loop chain with large $L_{c,B}$ must have considerable entanglements or kinks with other chains; otherwise it would have a shorter $L_{c,B}$. A tie chain can have very few kinks or entanglements and still have a large $L_{c,B}$ due to the fact that it has to span from one glassy domain to another.  The increased interactions or entanglements with other chains most likely increase the stress that chain is experiencing. Other geometrical measures do not show a clear correlation with stress.

\section{References}
\bibliography{bibliography}